\documentclass[11pt]{article}
\usepackage[utf8]{inputenc}

\usepackage[table]{xcolor}
\usepackage{authblk}
\usepackage{natbib}
\usepackage{graphicx}
\usepackage{multicol}
\usepackage[hidelinks]{hyperref}
\usepackage{caption} 
\usepackage{booktabs}
\usepackage{algorithm}
\usepackage{algorithmic}

\usepackage{cleveref}
\graphicspath{{Fig/}}
\usepackage{xcolor}
\usepackage{graphicx}
\usepackage[caption=false]{subfig}
\usepackage{tabularx} 

\usepackage{booktabs} 
\usepackage{rotating}
\usepackage{nth}
\usepackage{tikz}
\usepackage{multirow}
\usepackage{makecell}

\usepackage{geometry}
\date{}

\title{\textbf{Deterministic access to global viral sequence data enables robust agentic scientific discovery}}

\author[1,2$\ast$]{Ferdous Nasri}
\author[2,3,4]{Sarah Gurev}
\author[2]{Patrick Varilly}
\author[5]{Krithik Ramesh}
\author[6]{Nuala A. O'Leary}
\author[7]{Jonah Cool}
\author[1]{Bernhard Y. Renard}
\author[2,8,9,10$\ast$]{Pardis C. Sabeti}
\author[2,4,9$\ast$]{Laura Luebbert}

\affil[1]{Hasso Plattner Institute, Digital Engineering Faculty, University of Potsdam, Potsdam, Germany}
\affil[2]{Broad Institute of MIT and Harvard, Cambridge, MA, USA}
\affil[3]{Department of Biology, Massachusetts Institute of Technology, Cambridge, MA, USA}
\affil[4]{FutureHouse, San Francisco, CA, USA}
\affil[5]{Lyra Labs, Cambridge, MA, USA}
\affil[6]{National Center for Biotechnology Information, National Library of Medicine, National Institutes of Health, Bethesda, MD, USA}
\affil[7]{Anthropic, San Francisco, CA, USA}
\affil[8]{Department of Immunology and Infectious Diseases, Harvard T H Chan School of Public Health, Boston, MA, USA}
\affil[9]{Department of Organismic and Evolutionary Biology, Harvard University, Cambridge, MA, USA}
\affil[10]{Howard Hughes Medical Institute, Chevy Chase, MD, USA}
\affil[$\ast$]{Corresponding authors. 
\href{mailto:Ferdous.Nasri@hpi.de}{ferdous.nasri@hpi.de}, \href{mailto:Pardis@broadinstitute.org}{pardis@broadinstitute.org}, \href{mailto:luebbert@broadinstitute.org}{luebbert@broadinstitute.org}}

\begin{document}
\maketitle

\abstract{
Public viral genome resources such as the National Center for Biotechnology Information (NCBI) Virus database are central to outbreak response, evolutionary analysis, vaccine design, and genomic surveillance. However, many high-value retrieval workflows remain optimized for interactive use rather than exposed through deterministic, reproducible programmatic interfaces. This creates a particular challenge for Large Language Model (LLM)-based scientific agents, where small errors in metadata interpretation, filtering logic, or sequence retrieval can propagate into incorrect downstream datasets. To systematically evaluate agentic viral data retrieval, we built VirBench, a manually curated benchmark of 120 queries spanning diverse pathogens, taxonomic levels, and combinatorial metadata filters reflective of real-world virology workflows. When state-of-the-art autonomous AI systems, including Biomni, Claude, GPT, and Edison Analysis, were tasked with these queries without a dedicated retrieval layer, performance varied widely: mean accuracy ranged from 16.9\% for Claude Sonnet 4 to 91.3\% for GPT-5.5, with newer frontier models showing substantial progress but residual errors remaining consequential for reproducible dataset construction. To address this limitation, we built \textit{gget virus}, a deterministic query framework that formalizes and extends NCBI Virus-style filtering as a reproducible programmatic system. By staging retrieval, applying metadata constraints before sequence download, and selectively retrieving structured GenBank records, \textit{gget virus} reduces data transfer by more than 98\% for representative high-volume queries while preserving exact-match semantics. Instructing autonomous AI systems to use \textit{gget virus} increased accuracy to at least 90.0\% across all evaluated systems and up to 99.7\% for GPT-5.5, improved response stability to 0.92--1.00, reduced error magnitude, and generally decreased runtime and tool calls. Together, this work establishes deterministic data access as critical infrastructure for reliable agentic science and provides a reproducible retrieval layer for robust human- and AI-driven viral genomics workflows.

}

\section*{Introduction}
Advances in virology, immunology, and pathogen genomics increasingly depend on the ability to systematically retrieve, filter, and curate viral sequence data at scale. From vaccine antigen design and therapeutic antibody discovery to viral surveillance and host–pathogen modeling, nearly every downstream computational workflow begins with a query to a public viral sequence repository. The quality, reproducibility, and completeness of these initial queries directly determine the reliability of the resulting models and biological conclusions.

Despite this foundational role, programmatic access to viral sequence data remains underdeveloped. Existing interfaces are often incomplete, non-deterministic, and lack mechanisms to verify whether retrieved datasets fully satisfy query constraints. These limitations are particularly consequential in automated workflows, where errors in data retrieval cannot be easily detected or corrected.

The National Center for Biotechnology Information (NCBI) Virus resource \cite{ncbi, ncbivirus} is the primary public portal for viral sequence data, aggregating globally synchronized records across multiple underlying databases. NCBI Virus compiles and organizes viral records derived from GenBank® (all submitted records) and RefSeq (annotated reference sequence records), BioSample, Sequence Read Archive (SRA), and Assembly databases, which are federated through the International Nucleotide Sequence Database Collaboration (INSDC) \cite{insdc}. The INSDC is a long-standing global collaboration that synchronizes nucleotide sequence data across NCBI (USA), the European Nucleotide Archive (ENA) at EMBL-EBI (Europe) \cite{ena}, and the DNA Data Bank of Japan (DDBJ) (Japan) \cite{ddbj}, collectively representing the largest globally synchronized repository of openly accessible nucleotide sequence data. 

Despite rich web-based filtering capabilities, the NCBI Virus portal does not provide a comprehensive programmatic interface for reproducible data retrieval. While the web interface enables filtering by taxonomy, host, geographic location, sequence length, completeness, collection date, and other metadata, these capabilities are not fully exposed through programmatic access. This limitation arises in part because NCBI Virus is not a database itself, but a portal aggregating multiple underlying resources. The available NCBI APIs, including the NCBI Datasets REST API, NCBI Datasets CLI, and E-utilities (Entrez API), provide only a constrained subset of filtering functionality and do not fully expose the structured query capabilities available through the web portal~\Cref{tab:ncbi_apis}. As a result, researchers often resort to downloading large, minimally filtered datasets and performing extensive post hoc filtering, or following informal, non-versioned navigation workflows. Critically, these approaches provide no guarantees of completeness, making it difficult to determine whether retrieved datasets fully satisfy the intended query constraints. These approaches are inefficient, difficult to reproduce, and prone to inconsistencies across runs and users, resulting in non-deterministic dataset construction and limiting the reliability of downstream analyses.

As viral sequence datasets and automated analysis pipelines continue to scale, the absence of a robust and fully featured programmatic interface has become a critical bottleneck. The scale of viral sequence data has expanded dramatically, particularly following the SARS-CoV-2 pandemic, where millions of sequences were collected. Simultaneously, the research community increasingly relies on automated pipelines and artificial intelligence (AI)-driven systems, including emerging autonomous scientific agents such as Biomni \cite{biomni} and Edison Scientific's Kosmos \cite{kosmos}, and the underlying large language models (LLMs) that enable them, including GPT and Claude, to construct training datasets, generate benchmarks, and evaluate candidate antigens or therapeutics. In such settings, small inconsistencies in filtering logic can propagate into substantial downstream variation in model behavior and biological interpretation. Reproducibility, already a challenge in large-scale data science, becomes especially fragile when the data retrieval layer itself is not formally specified or programmatically controlled.

To address the gap in viral sequence retrieval, we developed \textit{gget virus}, an open-source command-line and Python tool for deterministic, programmatic access to viral sequence data. Rather than wrapping an existing API, \textit{gget virus} provides a structured interface to filtering logic that previously existed only implicitly within the NCBI Virus web interface, enabling reproducible, scriptable dataset construction for large-scale computational workflows. By formalizing viral sequence querying as a first-class computational operation, this approach transforms a previously ad hoc and error-prone process into reproducible scientific infrastructure. We evaluated whether this type of deterministic data access improves the correctness, scalability, and reliability of viral sequence retrieval, particularly in the context of automated and AI-driven analyses. To this end, we constructed a benchmark of real-world retrieval tasks and assessed both traditional programmatic methods and state-of-the-art autonomous AI systems.

\section*{Results}

\subsection*{VirBench benchmark}

We define deterministic viral sequence retrieval as the property that a query~$q$, executed against a fixed database state~$D$, yields a uniquely defined set of accessions $R(q,D)$ that is complete, reproducible across repeated executions, and invariant to implementation details such as pagination, batching, or API response ordering. In this setting, completeness requires that all records satisfying the query constraints are included in~$R(q,D)$, while soundness requires that all returned records satisfy those constraints. These elements in conjunction ensure that dataset construction is verifiable and stable, enabling reliable downstream analysis.

To systematically investigate failure modes in deterministic viral sequence retrieval, particularly in the context of language model–driven workflows, we constructed VirBench, an expert-curated benchmark of 120 viral sequence queries with ground-truth counts established through manual execution in the NCBI Virus web interface. Rather than serving only as a performance benchmark, VirBench was designed to isolate specific challenges inherent to programmatic viral data access, including multi-constraint filtering and large-scale pagination.

To this end, we constructed queries with progressively increasing levels of complexity, spanning diverse virus families and combinations of metadata constraints to emulate realistic and increasingly difficult retrieval scenarios. This design enables controlled evaluation of how systems handle the combinatorial structure of biological queries, as well as their ability to correctly aggregate results across large and heterogeneous datasets. Queries reflect plausible scientific use cases, including 58 contributed by the Sabeti Lab diagnostics team based on their frequent use in diagnostic assay development for priority pathogens \cite{gopal2025rapid, stachler2025establishing}.

Queries span 40 pathogens across multiple taxonomic levels, from broad searches over all viruses or entire families (e.g., \textit{Flaviviridae}) to species-level queries and individual accession lookups. Queries specify a pathogen (via NCBI TaxID or accession ID), with the exception of two queries that search across all viruses, and a combination of up to 16 filters, including host, geographic location, collection and release date ranges, sequence length bounds, nucleotide completeness, source database, genome segment, and vaccine strain status. Each query includes between 1 and 9 simultaneously applied filters (median 6), which were intentionally constructed to reflect the heterogeneity and difficulty of real-world virological analyses, ranging from simple cases (few filters, common pathogens) to complex scenarios (many filters, rare pathogens, or edge cases such as zero expected results). Expected sequence counts range from 0 to 3,226 (median 22). To ensure temporal stability, all queries except three accession-based lookups include a maximum release date. These manually verified counts serve as a reference standard for evaluating viral sequence retrieval systems in realistic virology research settings.

\subsection*{Limitations of LLM agents in viral sequence querying}

We evaluated the performance of autonomous AI research agents on programmatic viral sequence retrieval tasks, a critical prerequisite for reliable genomic surveillance, outbreak response, and downstream analysis, using the VirBench benchmark. We tested six autonomous AI research agents: Biomni \cite{biomni} (Claude Sonnet 4 backend), Edison Analysis \cite{edison_analysis}, GPT-5.2-pro (with web search and code execution tools), GPT-5.5 (with web search and code execution tools), Claude Sonnet 4 (with web search and code execution tools), and Claude Opus 4.7 (with web search and code execution tools). Claude Sonnet 4 represents the latest publicly available Anthropic model that can be used for this evaluation, due to subsequent biosafety-related access restrictions on newer models. Agents were provided with structured natural-language prompts constructed from each VirBench query. Agents were asked to return the final count of sequences matching the specified criteria, and each prompt was repeated three times.

Example prompt:
\texttt{"Retrieve viral sequences from NCBI for TaxID 3052310 (Lassa virus (LASV)) with the following filters: geographic location: Africa, collection date from 2020-01-01,\linebreak collection date until 2025-12-31. Return the count of sequences that match these criteria."}

\begin{figure*}[t]
    \centering
    \includegraphics[width=\linewidth]{images/agent_heatmap.pdf}
    \caption{\footnotesize
    \textbf{Agent performance and failure modes across VirBench taxonomies.}
    VirBench consists of 120 manually curated viral data retrieval queries, each evaluated across three independent runs per agent.
    \textbf{(A)} Mean accuracy per taxonomic target for each agent without (top) and with (middle) \textit{gget virus}, as well as the accuracy delta with versus without \textit{gget virus} (bottom). n denotes the number of queries covering the taxonomic target. Each query is repeated three times.
    \textbf{(B)} Distribution of deviations (retrieved minus expected sequence counts) for each agent, shown without (top) and with \textit{gget virus} (bottom). Bar plots show the total number of queries that returned more sequences than expected (over) and fewer sequences than expected (under).
    \textbf{(C)} Outcome breakdown across runs, showing the proportion of correct results, correct zero results, incorrect zero results, under-counting, over-counting, missing results, and errors for each agent with and without \textit{gget virus}.
    }
    \label{fig:agent_heatmap}
\end{figure*}

\begin{figure*}[t]
    \centering
    \includegraphics[width=\textwidth]{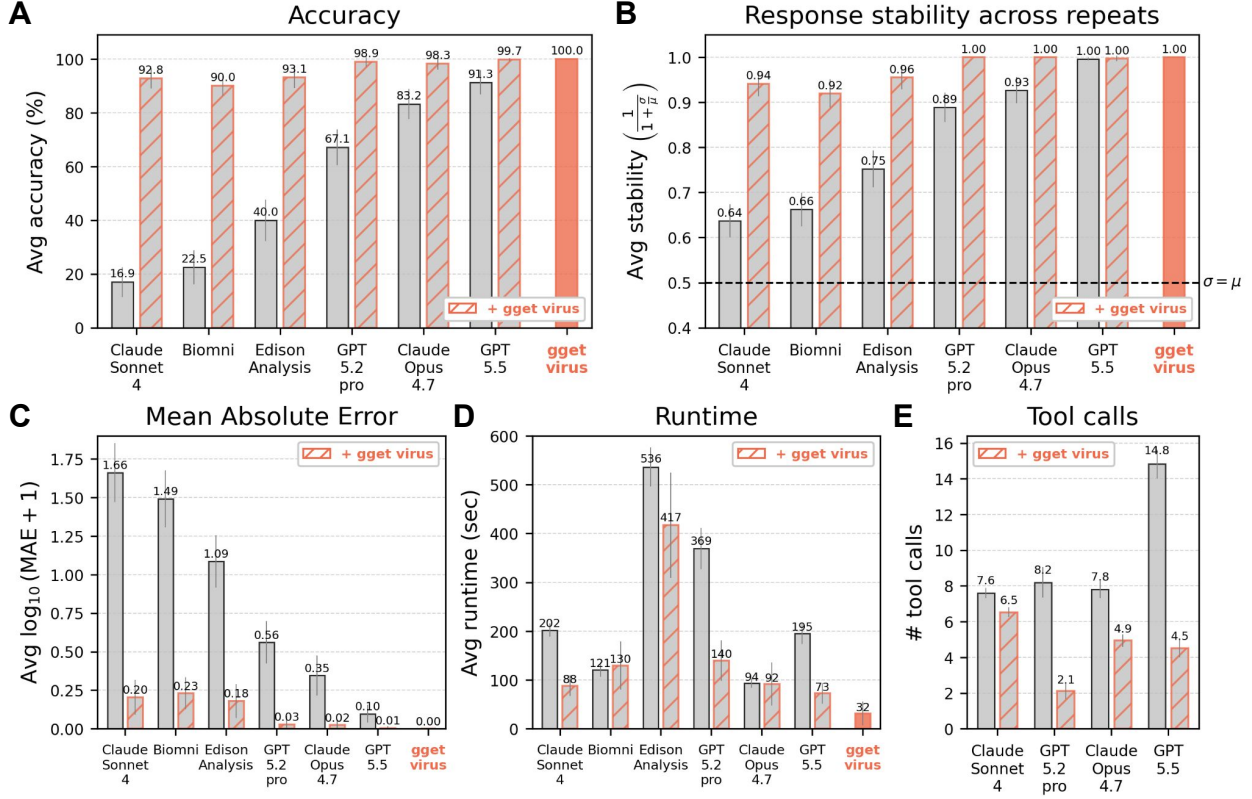}
    \caption{\footnotesize
    \textbf{Aggregate agent performance on VirBench with and without \textit{gget virus}.} Full per-query distributions are shown in \Cref{fig:supp_agents}. Results are shown for direct querying of \textit{gget virus}, and six AI research agents with and without being prompted to use \textit{gget virus}. Bars show the mean across queries with 95\% confidence interval (CI) error bars.
    \textbf{(A)} Accuracy, defined per query as the fraction of runs (out of three) that exactly match the expected sequence count.
    \textbf{(B)} Stability across repeated runs, computed per query as the bounded coefficient of variation ($1/(1+\sigma/\mu)$), where $\mu$ and $\sigma$ denote the mean and standard deviation of retrieved counts across runs. The dashed line marks the condition $\sigma = \mu$, corresponding to major instability across repeated runs.
    \textbf{(C)} Error magnitude measured as $\log_{10}(\mathrm{MAE}+1)$, where MAE is the mean absolute error between retrieved and expected counts per query across runs.
    \textbf{(D)} Runtime per query, averaged across runs.
    \textbf{(E)} Number of agent tool calls per query, averaged across runs.
    }
    \label{fig:agents}
\end{figure*}

When tasked with routine viral data retrieval, agent performance varied widely across systems and improved substantially in newer frontier models, supporting VirBench as a benchmark that captures meaningful progress in agent capability (\Cref{fig:agents}). However, even the strongest models did not consistently achieve the level of accuracy and reproducibility required for reliable dataset construction. Mean accuracy was 16.9\% for Claude Sonnet 4, 22.5\% for Biomni, 40.0\% for Edison Analysis, 67.1\% for GPT-5.2-pro, 83.2\% for Claude Opus 4.7, and 91.3\% for GPT-5.5\footnote{In one of 360 runs (Q32, run 3), GPT-5.5 independently identified and used \textit{gget virus}, despite not being explicitly prompted to do so. This was the only run for this question that produced the correct answer.} (\Cref{fig:agents}A). Stability across repeated runs also improved with newer models, reaching 0.93 for Claude Opus 4.7 and 1.00 for GPT-5.5, but earlier systems showed substantial variability when identical queries were submitted repeatedly (\Cref{fig:agents}B). Errors were frequently large in magnitude for earlier models, with log$_{10}$(MAE+1) values of 1.66 for Claude Sonnet 4, 1.49 for Biomni, 1.09 for Edison Analysis, and 0.56 for GPT-5.2-pro (\Cref{fig:agents}C), reflecting many instances of incorrect or incomplete retrieval. Even for higher-performing frontier models, residual errors remain consequential because a small number of incorrect retrievals can propagate into downstream analyses.

This issue is particularly consequential because viral sequence retrieval is typically an early step in a longer research workflow, such as phylogenetic inference, variant tracking, vaccine target selection, and machine learning model training. Agent-retrieved datasets often appear plausible even when incomplete or incorrectly filtered, making such errors difficult to detect through manual inspection of downstream outputs. Inaccuracies at the data retrieval stage can silently propagate through subsequent analyses, affecting model and benchmarking behaviors as well as biological interpretations. Deterministic and reproducible data access is therefore not merely a technical convenience, but a prerequisite for reliable automated scientific workflows.

When the same agents were instructed to use the \textit{gget virus} tool, described in the next section, for viral data retrieval, performance improved substantially and became more consistent across systems. Accuracy increased to 92.8\% for Claude Sonnet 4, 90.0\% for Biomni, 93.1\% for Edison Analysis, 98.9\% for GPT-5.2-pro, 98.3\% for Claude Opus 4.7, and 99.7\% for GPT-5.5 (\Cref{fig:agents}A), with high stability across repeated runs (0.92--1.00; \Cref{fig:agents}B). Mean absolute error was reduced to log$_{10}$(MAE+1) values of 0.18--0.23 for earlier systems and 0.01--0.03 for frontier models (\Cref{fig:agents}C). Using \textit{gget virus} also generally reduced both runtime and the number of tool calls required to complete each task. The largest runtime reductions were observed for Claude Sonnet 4, Edison Analysis, GPT-5.2-pro, and GPT-5.5, while the largest reductions in tool calls were observed for GPT-5.2-pro, Claude Opus 4.7, and GPT-5.5 (\Cref{fig:agents}D,E).

In the absence of \textit{gget virus}, performance varied substantially across viruses and query types, revealing failure modes in agent-only retrieval. The lowest accuracies were observed for, amongst others, Marburg virus, Influenza A virus, and queries searching all viruses (\Cref{fig:agent_heatmap}A), with Influenza A, Marburg, Mpox, and HIV-1 queries showing the highest deviations from the expected result (\Cref{fig:all_deviations}). For large datasets such as Influenza A ($>$1.5 million records) and SARS-CoV-2 ($>$9 million records), NCBI provides cached downloads that \textit{gget virus} leverages where available, avoiding page-by-page retrieval via the Datasets REST API. In contrast, agents that default to the API must retrieve data sequentially in batches of up to 1,000 records, a process that is slow and prone to failure, often terminating partway through multi-batch downloads with no straightforward way to resume. In \textit{gget virus}, these issues are mitigated by detecting failure patterns and retrying requests with adjusted batch sizes. Additionally, transient API issues can cause specific filters to fail for certain taxon IDs, including Mpox and HIV-1. \textit{gget virus} addresses this by retrying with alternative filter combinations and, when necessary, falling back to the highest level of accessible metadata.

Agents also struggled with specific filter types, particularly those requiring flexible string matching, such as segment and submitter country (\Cref{fig:filter_heatmaps}A). In addition to variability in how metadata fields are populated, the location of relevant information within records can also differ. For example, one VirBench query requests Influenza A sequences from Korea with additional filters. Although three sequences matching all filters exist, their geographic information is not stored in the expected location fields: the \texttt{location} field is empty, and the \texttt{region} field is recorded as "Asia." Instead, the relevant geographic signal appears in the virus name itself (e.g., "A/Korea/426/1968(H2N2)"). To account for such inconsistencies, the \texttt{geographic\_location} filter in \textit{gget virus} searches across multiple fields, including \texttt{location}, \texttt{region}, and \texttt{virusName}, emulating the behavior of the NCBI Virus web interface.

More broadly, error rates increased with query complexity, with performance degrading sharply beyond three to four simultaneous filters for agent-only approaches (\Cref{fig:filter_heatmaps}B). Across queries, agents under-counted sequences, consistent with incomplete pagination, premature termination of retrieval, or failure to aggregate results across multiple pages. For some agents, over-counting occurred more frequently than under-counting, indicating systematic issues in applying filtering criteria correctly (\Cref{fig:agent_heatmap}B, C).

These issues were largely resolved when agents were instructed to use \textit{gget virus}, which consistently improved accuracy and reduced both under- and over-counting across viruses and filter conditions (\Cref{fig:agent_heatmap}A; \Cref{fig:filter_heatmaps}A). Remaining errors generally reflected failures in how agents used or post-processed the tool output, rather than failures of the retrieval layer itself. The most common residual failure mode was incorrect local filtering after download, for example when agents reapplied metadata or sequence-level filters incorrectly to otherwise valid \textit{gget virus} results. Other errors occurred when agents partially processed large FASTA files, ignored the instruction to use \textit{gget virus} and reverted to alternative APIs (2 runs for Claude Sonnet 4 and 25 runs across 24 queries for GPT-5.2-pro), or called \textit{gget virus} with incorrect parameters despite iterative revisions to make the documentation more explicit, structured, and agent-accessible. Thus, the remaining errors were no longer primarily due to the absence of a deterministic retrieval interface, but to the agent’s ability to invoke that interface correctly and preserve its outputs without unnecessary reinterpretation.

Together, these results show that deterministic programmatic interfaces directly improve the reliability of autonomous scientific agents. Constraining agents to use \textit{gget virus} transformed viral data retrieval from a variable, error-prone process into a substantially more accurate and reproducible operation. At the same time, the residual failures underscore a second requirement for robust agentic science: tools must not only expose reliable data-access operations, but also return outputs and documentation that agents can use correctly without brittle post hoc processing.

\subsection*{Streamlined retrieval of large-scale viral datasets with \textit{gget virus}}

\begin{figure*}[t]
    \centering
    \includegraphics[width=1\linewidth]{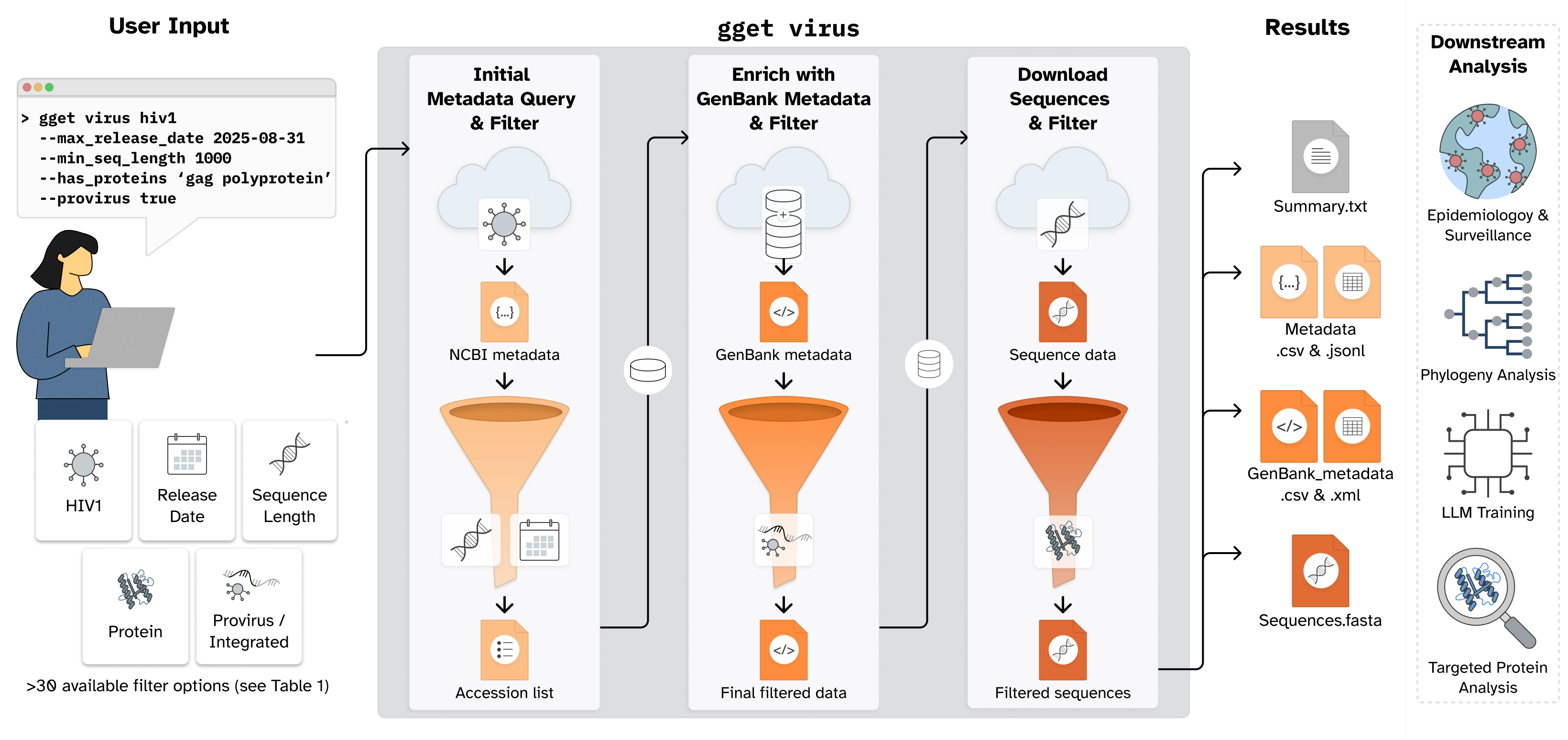} \\ 
    \caption{\footnotesize
    \textbf{Schematic overview of the \textit{gget virus} workflow} showing an exemplary search for proviral HIV1 sequences released before August 31st, 2025, with a minimum sequence length of 1,000 bases, and including the "gag polyprotein". The internal query processing steps are illustrated in the gray box, and the resulting output files returned to the user are shown on the right. A more detailed overview of the technical architecture of \textit{gget virus} is provided in \Cref{fig:overview2}.}
    \label{fig:overview}
\end{figure*}

\textit{gget virus} was designed to provide a structured, reproducible interface to NCBI Virus-style querying, which is challenging because the required information is distributed across multiple underlying resources rather than exposed through a single comprehensive programmatic endpoint. It therefore coordinates several NCBI access paths, including lightweight metadata retrieval through the NCBI Datasets REST API, cached bulk datasets through the Datasets command-line interface (CLI) for large resources such as SARS-CoV-2 and Influenza A, and detailed feature annotations through GenBank records retrieved via E-utilities (\Cref{fig:overview}). The system exposes the full range of filtering capabilities available through the web portal, including taxonomy, host, geographic location, gene, segment, sequence completeness, and additional metadata constraints, while extending these capabilities through structured filtering of GenBank-derived annotations such as proteins, genotype specifications, isolation source, and environmental context (\Cref{tab:feature-comp}). Results are returned in both human-readable and machine-readable formats. By automating access that previously required navigating complex web interfaces or fragmented API workflows, \textit{gget virus} enables reproducible, scriptable dataset construction suitable for large-scale computational analyses.


\textit{gget virus} implements a staged query optimization strategy that minimizes data transfer while preserving deterministic retrieval semantics. To avoid the creation of large temporary datasets, \textit{gget virus} employs an intelligent query optimization strategy and orders its operations to minimize data transfers, abstracting away the complexity of URL construction, pagination handling, accession batching, retry logic, and JSON/XML parsing. The different paths are illustrated in \Cref{fig:overview2}. \textit{gget virus} selects and optimizes the appropriate processing pathway based on the user’s query. In the exemplary case shown in \Cref{fig:overview}, \textit{gget virus} first downloads lightweight metadata while applying all available server-side filters, followed by local metadata filtering to construct a minimal accession set. Only the GenBank records corresponding to this filtered accession list are retrieved, and sequences are downloaded exclusively for records that pass all prior filters.

\begin{table*}[!t]
\centering
\small
\caption{Feature comparison of methods for viral sequence data retrieval and filtering.}
\label{tab:feature-comp}
\begin{tabularx}{\linewidth}{p{0.39\linewidth}|p{0.13\linewidth} X p{0.12\linewidth} p{0.1\linewidth}} 
\hline
\textbf{Feature}                                      & \textbf{NCBI Web} & \textbf{NCBI Datasets REST API} & \textbf{NCBI Datasets CLI}  & \textbf{\textit{gget virus}} \\ \hline
\multicolumn{4}{l}{\textit{General Capabilities}} \\
Scalability                                           & Low               & \textbf{High}                     & \textbf{High}          & \textbf{High}       \\
Optimised download pathway selection                  & N/A               & Manual                            & Manual                 & \textbf{Automatic}  \\
Chunked downloads for large datasets                  & N/A               & N/A                               & N/A                    & \textbf{Yes}        \\ 
New record / database update detection                  & N/A               & N/A                               & N/A                    & \textbf{Yes}        \\ \hline
\multicolumn{4}{l}{\textit{Basic Filtering Options}} \\
Virus name/taxon ID/accession                         & \textbf{Yes}      & \textbf{Yes}                      & \textbf{Yes}           & \textbf{Yes}        \\
Host specification                                    & \textbf{Yes}      & \textbf{Yes}                      & \textbf{Yes}           & \textbf{Yes}        \\
Geographic location                                   & \textbf{Yes}      & \textbf{Yes}                      & \textbf{Yes}           & \textbf{Yes}        \\
Pango lineage (for SARS-CoV-2)                        & \textbf{Yes}      & \textbf{Yes}                      & \textbf{Yes}           & \textbf{Yes}        \\
Minimum release date                                  & \textbf{Yes}      & \textbf{Yes}                      & \textbf{Yes}           & \textbf{Yes}        \\ \hline
\multicolumn{4}{l}{\textit{Advanced Filtering Options}} \\
Annotation status                                     & No                & \textbf{Yes}                      & \textbf{Yes}           & \textbf{Yes}        \\
Environmental source specification & Boolean only      & No                    & No & \textbf{Yes}        \\
Nucleotide completeness (complete or partial)         & \textbf{Yes}      & Only complete                     & Only complete & \textbf{Yes}        \\
Source database specification                             & \textbf{Yes}      & RefSeq only                    & RefSeq only           & \textbf{Yes}        \\
Assembly completeness                                 & \textbf{Yes}      & No                     & No & No\textsuperscript{$\ast$}       \\
Maximum release date                                  & \textbf{Yes}      & No                                & No                     & \textbf{Yes}        \\
Minimum sequence length                               & \textbf{Yes}      & No                                & No                     & \textbf{Yes}        \\
Maximum sequence length                               & \textbf{Yes}      & No                                & No                     & \textbf{Yes}        \\
Minimum collection date                               & \textbf{Yes}      & No                                & No                     & \textbf{Yes}        \\
Maximum collection date                               & \textbf{Yes}      & No                                & No                     & \textbf{Yes}        \\
Submitter name                                            & \textbf{Yes}      & No                                & No                     & \textbf{Yes}        \\ 
Submitter institution                                            & \textbf{Yes}      & No                                & No                     & \textbf{Yes}        \\
Submitter country                                     & \textbf{Yes}      & No                                & No                     & \textbf{Yes}        \\
Protein/gene specification                            & \textbf{Yes}      & No                                & No                     & \textbf{Yes}        \\
Lab passaged information                              & \textbf{Yes}      & No                                & No                     & \textbf{Yes}        \\
Maximum ambiguous characters                          & \textbf{Yes}      & No                                & No                     & \textbf{Yes}        \\
Segment specification                                 & \textbf{Yes}      & No                                & No                     & \textbf{Yes}        \\
Vaccine strain specification                          & \textbf{Yes}      & No                                & No                     & \textbf{Yes}        \\
Provirus or integrated specification                  & \textbf{Yes}      & No                                & No                     & \textbf{Yes}        \\
Isolate / Strain specification                                             & \textbf{Yes}      & No                                & No                     & \textbf{Yes}        \\
Genotype specification                                            & \textbf{Yes}      & No                                & No                     & \textbf{Yes}        \\ 
Tissue / Specimen / Source specification                                            & \textbf{Yes}      & No                                & No                     & \textbf{Yes}        \\  
Genome molecule type specification                                            & \textbf{Yes}      & No                                & No                     & \textbf{Yes}        \\ \hline
\multicolumn{4}{l}{\textit{Specialized Filtering (unique to gget virus)}} \\
Minimum gene count                                    & No                & No                                & No                     & \textbf{Yes}        \\
Maximum gene count                                    & No                & No                                & No                     & \textbf{Yes}        \\
Minimum protein count                                 & No                & No                                & No                     & \textbf{Yes}        \\
Maximum protein count                                 & No                & No                                & No                     & \textbf{Yes}        \\
Minimum Mature peptide count                          & No                & No                                & No                     & \textbf{Yes}        \\
Maximum Mature peptide count                          & No                & No                                & No                     & \textbf{Yes}        \\
Protein completeness                                  & No                & No                                & No                     & \textbf{Yes}        \\ \hline
\multicolumn{4}{l}{\textit{Output Options}} \\
Detailed GenBank metadata export                               & Manual            & Manual                            & Manual                 & \textbf{Built-in}   \\ \hline
\end{tabularx}
\caption*{\footnotesize $\ast$ Not exposed through any available API, and the underlying computation could not be inferred from the accessible metadata.}
\end{table*}

To enable efficient and flexible data retrieval, \textit{gget virus} leverages multiple NCBI interfaces, each suited to different stages of the workflow. Lightweight metadata is retrieved using the NCBI REST API, which supports server-side filtering but requires pagination for datasets exceeding 1,000 records. For specific viruses such as Alphainfluenza A and SARS-CoV-2, precomputed datasets with limited filtering options are available via the NCBI Datasets CLI. More detailed GenBank metadata and sequence data are accessed through the Entrez system using E-utilities. \textit{gget virus} integrates these APIs strategically, selecting the most appropriate data source at each stage to ensure efficient and comprehensive retrieval (\Cref{tab:ncbi_apis}). In addition, the workflow incorporates robust failure handling to account for inconsistencies across NCBI services. For example, when certain taxonomies transiently fail to support specific filters via the Datasets REST API (a behavior observed for a query described in \Cref{tab:process_comparison} at the time of writing this paper), \textit{gget virus} detects the error, and iteratively removes unsupported filters from the query, and reapplies them locally during downstream processing.

\begin{figure*}[t]
    \centering
    \includegraphics[width=1\linewidth]{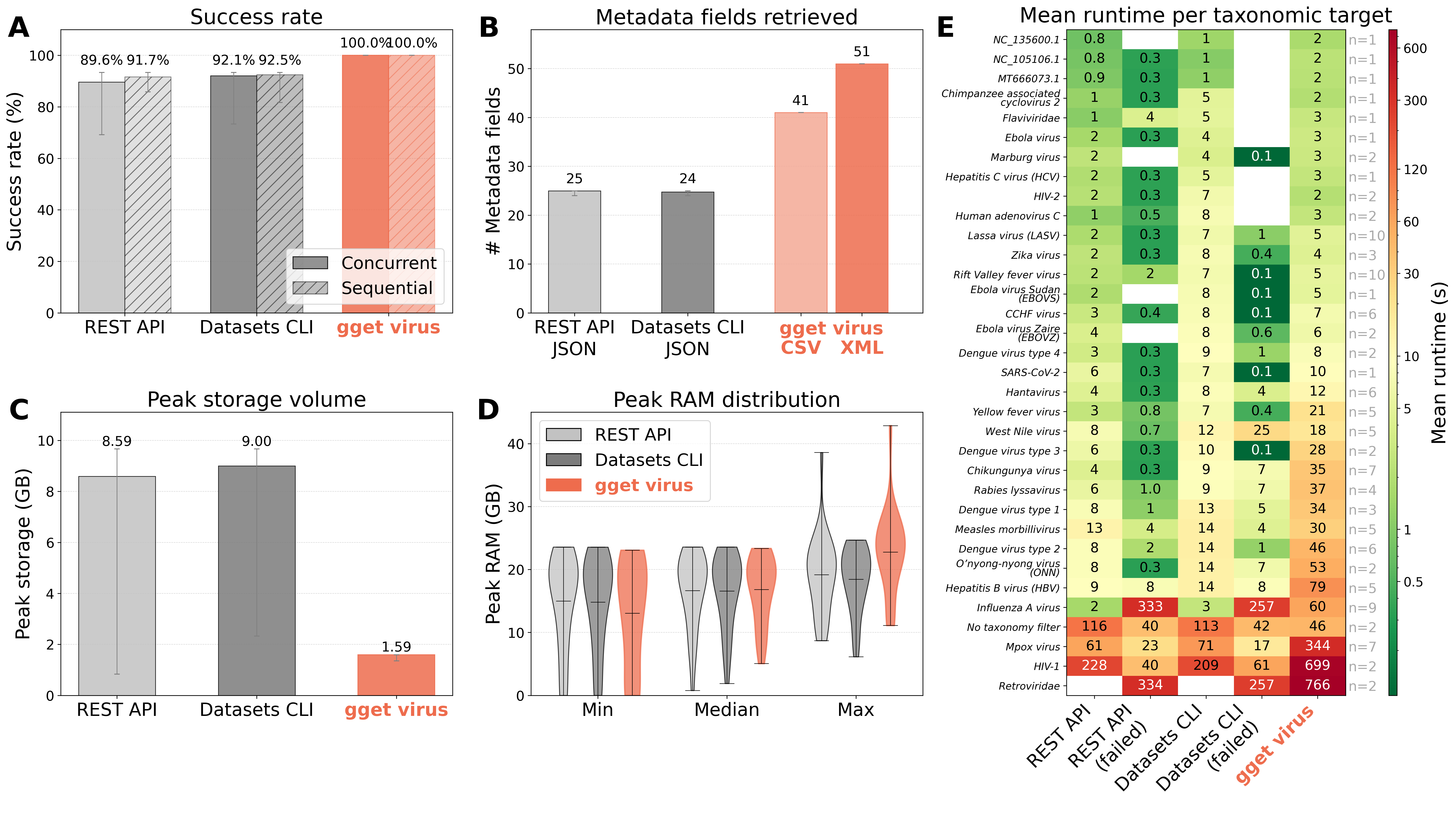}
    \caption{\footnotesize \textbf{Stress-testing programmatic viral data retrieval (120 queries, executed three at a time concurrently every hour over 24 hours)}. Results are shown for direct querying via the NCBI Datasets REST API, the NCBI Datasets CLI, and \textit{gget virus}.
    \textbf{(A)} Stress-test success rate, defined as the percentage of queries completed without errors. Bars represent the median success rates across rounds, and error bars represent the minimum and maximum observed rates. The results are shown for both concurrent (three simultaneous queries per method) and sequential (one query at a time) tests.
    \textbf{(B)} Bar plot showing the number of metadata fields retrieved per method and output format. \textit{gget virus} provides two enriched metadata output files, including NCBI (CSV) and GenBank (XML) data, which are shown separately. Bars show the median grouped per round, with error bars extending to the minimum and maximum across rounds.
    \textbf{(C)} Storage volume consumed per query (GB). Bars represent median data size stored per round over all queries, with error bars spanning the minimum and maximum round-level median.
    \textbf{(D)} Distribution of peak RAM usage per query (GB), shown as violin plots grouped by minimum, median, and maximum RAM computed within 120-query rounds. Each violin displays the spread across all rounds with means and medians marked, revealing both central tendency and round-to-round variability of minimum, median (typical), and maximum (worst-case) memory consumptions of these methods based on VirBench queries run concurrently at the defined API limit. Detailed per-query RAM usage for a sequential run is shown in \Cref{fig:ram-usage}.
    \textbf{(E)} Mean runtime per pathogen for each method (separately including failed runs for each method). \textit{n} denotes the number of queries covering the virus. White cells show the lack of runs with that status (successful / failed) over all repetitions. Failed runs for \textit{gget virus} are not shown since all runs were successful. Mean runtime for each query for sequential and concurrent benchmarking runs is shown in \Cref{fig:runtime_heatmaps}. Runtime distribution and throughput details shown in  \Cref{fig:runtime-dist-throughput}.
    }
    \label{fig:benchmark-methods}
\end{figure*}

The order of operations is optimized based on the requested filters. For example, when GenBank-level filters are specified, GenBank records are retrieved prior to sequence download; otherwise, GenBank data are fetched only after sequence-level filtering has been applied. This staged architecture avoids unnecessary transfer of full nucleotide sequences and large GenBank XML records, thereby reducing bandwidth usage, storage requirements, and memory consumption while maintaining full reproducibility (\Cref{fig:benchmark-methods}C, D).

As an illustrative example of this hybrid retrieval strategy, we retrieved all SARS-CoV-2 sequences released in 2025 containing the surface glycoprotein. The standard “download-then-filter” approach requires downloading 284 GB of data prior to local filtering, requiring self-written scripts. In contrast, the \textit{gget virus} strategy provides the user with only 3.8 GB of targeted data for this query, which represents a 98\% reduction in data volume storage, making large-scale viral sequence download computationally feasible on standard local hardware (details in \Cref{tab:process_comparison}).

\subsection*{Stress-testing programmatic data retrieval}

To systematically evaluate the performance and resource efficiency of \textit{gget virus}, we used the VirBench benchmark to compare it against existing programmatic retrieval methods, including direct interaction with the NCBI Datasets REST API and the Datasets CLI. All 120 curated queries were executed hourly over a 24-hour period to capture variability in latency and server-side stability, using both sequential and concurrent (three simultaneous queries) execution modes. Because the NCBI REST API and Datasets CLI lack several filtering capabilities required by the benchmark queries, we supplemented these runs with a naïve post-download filtering step to approximate equivalent functionality.

A primary bottleneck in the "download-then-filter" paradigm in large-scale viral genomics is the storage required. \textit{gget virus} achieves a statistically significant reduction in storage volume used in comparison to other programmatic methods (\Cref{fig:benchmark-methods}C, \Cref{tab:stat_tests}). The median per-query storage is 0.4\,MB for \textit{gget virus}, versus 2.2\,MB  for the Datasets REST API, and 2.6\,MB for Datasets CLI. Aggregated across a full benchmark round, this amounts to a median total of 1.59 GB for \textit{gget virus} versus 8.59 GB for Datasets REST API and 9.00 GB for Datasets CLI, representing an approximate 5- to 6-fold reduction in size.

Despite the smaller storage footprint, \textit{gget virus} enables richer metadata retrieval. The enriched CSV output contained 41 metadata fields per query, and the enriched GenBank metadata contained 51 metadata fields, compared with 25 fields for both Datasets REST API and Datasets CLI JSON output files (\Cref{fig:benchmark-methods}B).

Critically, these gains in storage efficiency and metadata richness do not come at the expense of increased memory consumption (\Cref{fig:benchmark-methods}D). Peak RAM usage is comparable across all three methods (\Cref{tab:stat_tests}). Sequential benchmarking further confirmed that \textit{gget virus} operates efficiently on standard consumer hardware, maintaining a memory footprint well below 16 GB even when several queries are submitted concurrently (\Cref{fig:ram-usage}).

To assess reliability under realistic usage conditions, we executed all queries repeatedly in concurrent runs. A successful retrieval was defined as returning the expected number of accessions without errors or run failures. Native NCBI methods exhibited variability across run, with fluctuating success rates across rounds (\Cref{fig:benchmark-methods}A). In contrast, \textit{gget virus} produced consistently stable results by employing internal retry mechanisms and fallback strategies to mitigate transient API failures (\Cref{fig:overview2}). Although these robustness features can increase maximum runtime under conditions of server-side instability (\Cref{fig:benchmark-methods}E, \Cref{fig:runtime_heatmaps}, \Cref{fig:runtime-dist-throughput}), they maximize the correctness and completeness of the final dataset.

\section*{Practical applications}

Deterministic and programmatic viral sequence retrieval through \textit{gget virus} enables a range of downstream applications, including viral surveillance, phylogenetics, dynamics, and evolution, and helps inform public health responses and vaccine and therapeutic design. Crucially, \textit{gget virus} enables filtering capabilities that are not natively accessible through the existing APIs and are not scalable using the web interface alone. Additionally, GenBank metadata is not automatically available for integrated download, but \textit{gget virus} retrieves and associates metadata at sequence resolution.

Many computational analyses rely on large-scale viral sequence datasets with associated metadata to identify patterns in genotype distribution, host range, and temporal dynamics. Datasets curated from NCBI Virus can reveal phylogenetic relationships and haplotypes, enabling inference of evolutionary history within specific regions and time periods \cite{suchard2018bayesian}. Phylodynamic approaches integrate these sequence data with geographic information, serology, and host species data to identify emerging variants and inform public health strategies \cite{hadfield2018nextstrain}.

The integrated sequence-metadata retrieval capabilities of \textit{gget virus} are particularly valuable for outbreak investigation, where rapid and reproducible dataset construction is essential, enabling transmission chain tracking through genomic epidemiology and monitoring of drug resistance mutations across temporal and geographic scales \cite{grubaugh2019tracking,holmes2018pandemics,inzaule2023recommendations}. Furthermore, curated viral datasets support machine learning models that predict viral evolution and inform vaccine design \citep{gurev2025evaluating,thadani2023learning}. This streamlined access to filtered and annotated viral data accelerates hypothesis testing and comparative genomic analyses that would otherwise require extensive manual data processing and curation.

\section*{Discussion}

\textit{gget virus} formalizes viral sequence retrieval as a deterministic and reproducible computational operation, eliminating a long-standing trade-off between usability and functionality in accessing large-scale viral datasets. By replacing bulk downloads and local filtering with structured queries, it enables precise, reproducible viral datasets construction, suitable for surveillance pipelines, evolutionary analyses, and automated modeling systems.

The importance of reproducible programmatic data access is particularly evident in outbreak settings such as recently witnessed for the SARS-CoV-2 and Mpox viruses, where the absence of standardized and reproducible download protocols slowed data analysis and hindered coordinated, time-sensitive response efforts \cite{hodcroft2021want, mpoxradar, ling2022challenges, knyazev2022unlocking}. Researchers frequently had to download large datasets and apply manual post hoc filtering due to limited programmatic access to rich filtering functionality. These workflows were inefficient, difficult to document, and challenging to reproduce. In rapidly evolving outbreak settings, delays and inconsistencies in dataset assembly can materially affect surveillance, variant tracking, and downstream modeling.

Through integration into the widely adopted \textit{gget} ecosystem, \textit{gget virus} lowers barriers to community uptake and facilitates seamless incorporation into high-throughput pipelines and reproducible research environments. Anchoring retrieval to the synchronized INSDC repositories (GenBank, ENA, and DDBJ) further strengthens long-term data accessibility and reduces dependence on transient web interfaces.

Despite its advantages, \textit{gget virus} is not without limitations. First it depends on external database infrastructure and therefore inherits constraints of underlying APIs and network availability. Although retry mechanisms and fallback strategies mitigate transient failures, data retrieval ultimately depends on remote system availability. Second, performance may be reduced in low-bandwidth settings, where connectivity constraints may affect reliability. Third, the accuracy of retrieved datasets is bounded by the quality of underlying databases, which may be incomplete, inconsistent, or have erroneous annotations. Filtering based on metadata such as location, collection date, or gene annotations can therefore propagate upstream inconsistencies, and \textit{gget virus} cannot correct such errors.

The broader implications of programmatic viral data access also extend to biosecurity. \textit{gget virus} operates on publicly available data and does not enable access to restricted, select agent, or otherwise controlled biological information beyond what is already accessible through standard database interfaces. However, as with other tools that facilitate large-scale data access, its integration into automated or AI-driven pipelines raises potential dual-use considerations. The risks are comparable to those associated with existing database interfaces but may be amplified through automation \cite{bloomfield2024ai,wang2026without}. Balancing these considerations with the clear public health benefits of improved data access will be important as such tools are adopted more widely.

Finally, our evaluation of autonomous AI research agents highlights both the promise and the current limitations of automated scientific systems. When tasked with representative viral sequence retrieval queries, agents exhibited substantial variability and low baseline accuracy in the absence of structured programmatic access. Prompting agents to use \textit{gget virus} markedly improved both correctness and stability, demonstrating that reliable AI-driven workflows depend critically on deterministic data access. At the same time, these results represent a snapshot in time. Agent behavior will continue to evolve as models are updated and as external tools become more widely recognized and integrated into training and prompting strategies. It is conceivable that agents may increasingly default to structured retrieval tools such as \textit{gget virus}, reducing the observed performance gap. Nonetheless, the central finding remains: reproducible scientific automation requires stable, well-defined programmatic interfaces, independent of transient model behavior.

\section*{Methods}

\subsection*{\textit{gget virus} algorithm and implementation}

\subsubsection*{Usage and documentation}

\textit{gget virus} seamlessly integrates into the widely used \textit{gget} ecosystem \cite{gget}, adhering to the philosophy of "one-line" data retrieval and allowing users to query viral sequences directly from the command line or Python scripts. \textit{gget} can be installed using \textit{uv} or \textit{PyPI} via the command line:
\texttt{\$ uv pip install gget} or \texttt{\$ pip install --upgrade gget}.

Comprehensive documentation of all filters is available via the \texttt{--help}/\texttt{-h} flag on the command line: \texttt{\$ gget virus -h}, as function descriptions within the Python environment, or via the documentation website: \url{https://pachterlab.github.io/gget/en/virus.html} (also see \Cref{tab:links}). Software stability is ensured through extensive unit tests that automatically run twice a week. Users can access the latest unit test report here: \url{https://github.com/pachterlab/gget/blob/main/tests/pytest_results_py3.12.txt}.

Below is a simple example query for complete Zika virus genomes found in human hosts:

\begin{verbatim}
$ gget virus "Zika virus" \
--nuc_completeness complete \
--host human
\end{verbatim}

The tool can also be loaded as a Python module to allow dynamic pipeline integrations and automated reports on viral variants. The following code can be used to get the same data as above, already loaded into a Pandas DataFrame for immediate use:

\begin{verbatim}
import gget

# Fetch metadata for analysis
df = gget.virus(
    "Zika virus",
    nuc_completeness="complete",
    host="human"
)

# Ready for immediate downstream usage
print(f"Retrieved {len(df)} sequences.")

\end{verbatim}

\subsubsection*{Query optimization strategy}

\textit{gget virus} first queries the NCBI Datasets REST API \cite{covid-downloader} to download only lightweight metadata in an initial step. This first request downloads only dataset reports, applying all available server-side filters exposed by the API (e.g., host, release date, completeness) (\Cref{tab:feature-comp}). The remaining metadata filters are applied client-side to the downloaded metadata file (e.g., minimum/maximum sequence length, gene count, submitter country). Only after this step, a list of accession IDs is created and used to download the corresponding nucleotide sequences or to retrieve GenBank metadata records via \textit{E-utilities} (\Cref{fig:overview2}). This distinction is made based on whether GenBank-level filters are specified. Because GenBank records are substantially larger and more deeply nested (containing detailed information such as protein annotations, coding sequence boundaries, and amino acid sequences), they are only retrieved when required. When such filters are requested, \textit{gget virus} first downloads the relevant GenBank metadata and applies filters that depend on these annotations (e.g., environmental source or proviral state, referring to a dormant viral genome integrated into a host). Only after applying these filters, a refined list of accession IDs is generated for downstream sequence retrieval.

Next, \textit{gget virus} applies sequence-dependent filters (e.g., maximum ambiguous characters, presence of certain proteins). Subsequently, if requested (and not already done), the GenBank metadata is retrieved only for the sequences passing all previously applied filters. This staged filtering strategy ensures a logical application of filters and downloads, realizing the user’s query with minimal storage requirements.

Importantly, the initial metadata download contains only the fields necessary to implement metadata-level filtering and construct accession lists. The GenBank retrieval step later downloads the full structured record for the final accession set. This distinction prevents unnecessary transfer of large XML-formatted GenBank records for sequences that would ultimately be discarded.

\subsubsection*{Hybrid retrieval for large viral datasets}

For massive viral datasets, such as SARS-CoV-2 or Alphainfluenza (Influenza A), \textit{gget virus} automatically leverages highly compressed cached data packages via the \textit{NCBI Cached Datasets CLI} \cite{covid-downloader} to ensure speed and reliability. After a user sets the desired filters, \textit{gget virus} automatically selects the optimal method of data retrieval (\textit{NCBI Cached Datasets CLI} or direct \textit{NCBI Datasets REST API} requests). This distinction lowers the risk of overwhelming server entry points and increases the scalability of data queries.

\subsubsection*{Memory efficiency}

To keep memory usage independent of total dataset size, \textit{gget virus} processes data in configurable batches. Metadata retrieval is streamed and filtered incrementally, and sequence downloads are performed in accession batches. Sequence-level filtering is also performed in a streaming manner: each FASTA record is evaluated individually and, if it passes the specified filters, written directly to disk without accumulating filtered sequences in memory. Similarly, metadata output is written in chunks rather than assembled as a single monolithic in-memory object. Intermediate data structures are explicitly released between processing stages to further reduce peak memory usage. As a result, RAM requirements scale primarily with user-defined batch sizes rather than with the total number of viral sequences queried.

\textit{gget virus} also supports incremental dataset updates through baseline deduplication. Users may provide a metadata file from a previous run, in which case only novel accessions absent from the baseline are retrieved. This enables efficient resumption of interrupted downloads and incremental updates of local NCBI Virus datasets, without redundant data transfer.

\subsubsection*{Output structure}

Metadata are output in CSV and JSONL formats, while nested GenBank metadata are provided in CSV and XML formats for both human readability and programmatic downstream use. Nucleotide sequences are saved in FASTA files.

To further increase reproducibility, a text summary file is generated for each \textit{gget virus} query. This file records the exact parameters used, \textit{gget} version used, runtime, data transfer size, filtering statistics, and names of files produced, along with a brief report on the downloaded data. Failed API operations, if any, are logged together with exact retry URLs to ensure deterministic reconstruction of the dataset.



\subsection*{Stress-testing programmatic data retrieval}

To compare previously available programmatic methods to retrieve viral data from NCBI, we converted the 120 queries from VirBench into a structured JSON format to ensure identical parametrization across three tested methods: the NCBI Datasets REST API, NCBI Datasets CLI, and \textit{gget virus}. Benchmarks were executed both sequentially and concurrently over a 24-hour period, method by method. For concurrent runs, we restricted execution to three simultaneous queries per method, adhering to the most conservative rate limits specified in the NCBI API documentation to avoid artificial throttling \cite{covid-downloader}. 

During each execution, we recorded the following metrics: disk space occupied by the retrieved dataset (both sequences and metadata), RAM usage, success status of the run, and the number of unique metadata fields in the final output. RAM usage was sampled at 0.5-second intervals to determine instantaneous and peak memory consumption. A \textit{successful} run is defined as one completed without terminal errors and returning an accession count matching the manually curated VirBench ground truth. 
As the native NCBI tools do not support advanced filtering options, we implemented a simple script to further filter their outputs. This ensured comparability of resource usage required to reach a filtered dataset.

To evaluate the significance of the differences in storage consumption, metadata richness, and memory usage, we applied the two-sided Mann--Whitney $U$ test, a non-parametric test that compares the rank distributions of two independent samples without assuming normality or equal variances. Storage and metadata field counts are compared between \textit{gget virus} and each baseline method across all successful observations pooled over benchmark rounds. Metadata fields are compared per output format: CSV and XML from \textit{gget virus} are each tested against the JSON outputs of the Datasets REST API and Datasets CLI. All $p$-values are corrected for multiple comparisons using a Bonferroni adjustment over $k = 8$ tests, yielding a corrected significance threshold of
$\alpha_{\mathrm{adj}} = 0.05 / 8 = 0.00625$.

This benchmark was performed on an institutional network to account for bandwidth variables, while using publicly released \textit{gget virus} version 0.30.3, NCBI Datasets version 18.13.0, and NCBI Datasets REST API version 2.

\subsection*{AI agent benchmarking}

Each benchmark query was programmatically converted into a natural-language prompt using a shared build\_query() function, which translates structured filter columns into human-readable descriptions (e.g., "geographic location of sample collection: Africa; collection date on or after 2020-01-01; minimum sequence length: 11,000 bp"). These prompts were issued to four agentic systems: Edison Analysis (a proprietary agent accessed via the Edison client SDK and evaluated on February 26, 2026), Biomni v0.0.8 (a data-lake agent powered by Claude Sonnet 4), Claude Sonnet 4 (claude-sonnet-4-20250514, via the Anthropic Messages API), Claude Opus 4.7 (claude-opus-4-7, via the Anthropic Messages API), GPT-5.2-pro (via the OpenAI Responses API), and GPT-5.5 (via the OpenAI Responses API). Claude Sonnet 4 represents the latest publicly available Anthropic model that can be used for this evaluation, due to subsequent biosafety-related access restrictions on newer models. Each query was executed independently three times per agent to assess reproducibility.

\textbf{System prompt for Claude and GPT models:}
\texttt{You are a bioinformatics agent. You have access to: \texttt{web\_search}, which searches the internet for up-to-date information; and execute\_python, which runs Python locally with internet access to call APIs, install packages, and compute results. Outbound network access is restricted to an allowlist of domains. Workflow: first, use web\_search if needed to choose the best API or library; then use execute\_python to implement the call and compute the final answer. Output rules: when the final count is available, respond with exactly one integer on its own line. After outputting the final integer, also output a JSON block with keys methods and reasoning; the reasoning should contain at most three bullets and be kept brief.}

\textbf{Example prompt:}
\texttt{Retrieve viral sequences from NCBI for TaxID 3052310 (Lassa virus (LASV)) with the following filters: geographic location: Africa, collection date from 2020-01-01, collection date until 2025-12-31. Return the count of sequences that match these criteria.}

Claude and GPT models were operated in an agentic loop with up to 25 model turns and access to two tools: (i) server-side web search, used to identify relevant API documentation and usage examples, and (ii) local Python code execution, with a 120-second timeout per invocation. Outbound network access during code execution was restricted via a socket-level domain allowlist to NCBI (E-utilities, Datasets API, FTP) and PyPI domains, preventing data exfiltration to unauthorized endpoints. Models were instructed to return the final result as a single integer, which was parsed directly from their responses. Edison and Biomni managed tool use internally; their free-text outputs were converted into integer counts via a secondary parsing step using Claude Sonnet 4.

The \textit{gget virus} Python module (v0.30.3) was evaluated both as a standalone deterministic baseline, directly querying the NCBI Virus API without LLM involvement, and as an optional tool available to each agent. In this \textit{gget}-augmented setting, agents were instructed to install \textit{gget} by adding \texttt{"Use the gget virus module installable with 'pip install gget==0.30.3'."} to each prompt. Agents were moreover provided a slightly modified module documentation adapted for agent use (\url{https://github.com/lauraluebbert/VirBench/blob/main/docs/gget_virus_docs.md}), enabling use of \textit{gget virus} alongside or in place of alternative approaches discovered via web search.

Retrieved counts were compared against manually curated reference values to compute accuracy (exact match), run-to-run stability across the three independent executions per query, mean absolute error, and per-query runtime. Runs that resulted in explicit failures (e.g., timeouts, API errors, , or other agent-level execution failures) were rerun or excluded from metric calculations to minimize the influence of transient NCBI server instability, which is evaluated separately in dedicated stress tests. Runs that completed without producing a parseable count were also excluded. After exclusion of failed or unparseable runs, the analyses included 360/360 runs for Claude Sonnet 4, Claude Sonnet 4 + \textit{gget}, Biomni (Sonnet 4), and Biomni (Sonnet 4) + \textit{gget}; 355/360 for Edison Analysis; 349/360 for Edison Analysis + \textit{gget}; 357/360 for GPT-5.2-pro; 345/360 for GPT-5.2-pro + \textit{gget}; 353/360 for Claude Opus 4.7; 356/360 for Claude Opus 4.7 + \textit{gget}; 344/360 for GPT-5.5; 352/360 for GPT-5.5 + \textit{gget}; and 358/360 for \textit{gget virus}.

Additional details on the agent frameworks, prompts, and evaluation scripts are available at: \url{https://github.com/lauraluebbert/VirBench}

\section*{Code availability}
The manual and source code for \textit{gget virus} are publicly available under a BSD 2-Clause License at \url{https://pachterlab.github.io/gget/en/virus.html} and \url{https://github.com/pachterlab/gget/blob/main/gget/gget_virus.py}, respectively. Further links to documentation in Spanish, as well as Google Colab tutorials, are listed in \Cref{tab:links}. The code to reproduce the VirBench API and agent benchmarks can be found here: \url{https://github.com/lauraluebbert/VirBench}. 

The full VirBench benchmark dataset is not publicly released to prevent potential data leakage into large language model training corpora, which could compromise the validity of evaluation results by enabling models to retrieve or memorize answers. Researchers interested in accessing the benchmark are encouraged to contact the corresponding authors to request access.

\section*{Funding}
This work was supported in part by the National Center for Biotechnology Information of the National Library of Medicine (NLM), National Institutes of Health (NIH). The contributions of the NIH author(s) are considered Works of the United States Government. The findings and conclusions presented in this paper are those of the author(s) and do not necessarily reflect the views of the NIH or the U.S. Department of Health and Human Services. L.L. and S.G. are supported by funding from the FutureHouse AI-for-Science Postdoctoral Fellowship. L.L. is moreover supported by the Eric and Wendy Schmidt Center at the Broad Institute of MIT and Harvard. Work at FutureHouse and at the Eric and Wendy Schmidt Center is supported by the generosity of Eric and Wendy Schmidt. Furthermore, this work was supported by Anthropic through compute credits awarded to L.L.. F.N. and B.Y.R. are supported by the Deutsche Forschungsgemeinschaft (DFG) [459422098], and B.Y.R. is supported by a European Research Council (ERC) grant (eXplAInProt, 101124385). This work is made possible by support from the John D. and Catherine T. MacArthur Foundation, Flu Lab, and a cohort of generous donors through TED’s Audacious Project, including the ELMA Foundation, MacKenzie Scott, the Skoll Foundation, and Open Philanthropy. \textit{gget} is part of the scverse ecosystem (\url{https://scverse.org/}) and is supported by public donations through NumFOCUS.

\section*{Acknowledgements}
We thank Anthropic, especially Sreya Parakala, and OpenAI, especially Kenny Kim and Joy Jiao, for providing technical guidance and funding that facilitated the VirBench benchmarking runs. We thank the NCBI Virus development team, in particular Eneida Hatcher, for their responsive support and open communication during the development of \textit{gget virus}, and for providing an invaluable resource to the scientific community through NCBI Virus. We also acknowledge the funders, maintainers, and contributors of NCBI, ENA, and DDBJ for sustaining the international nucleotide sequence infrastructure and making it openly available. We thank the INSDC for enabling the routine synchronization and global accessibility of nucleotide sequence data across these repositories. We moreover thank Curtis Hoffmann for first drawing our attention to the challenges associated with viral sequence retrieval. We thank Nisha Gopal and Elyse Stachler for real-world viral sequence retrieval queries used in our benchmark, reflecting recurring assay development and surveillance workflows that motivated the automation capabilities of \textit{gget virus}. We thank Daniel J. Park and Cesar Arze for reviewing the manuscript and providing helpful comments, and Luciana Serna Wills for design support for Figures 3 and S5.



\section*{Competing interests}
P.C.S. holds several patents related to diagnostic technologies and is a cofounder and equity holder in Delve Biosciences and Lyra Labs, a board member and equity holder in Polaris Genomics, and an equity holder of NextGenJane. P.C.S. was formerly a co-founder of Sherlock Biosciences and board member of Danaher Corporation, until December 2024. B.Y.R. holds intellectual property rights commercialized by Seqstant, BioNTech, Genentech (Roche). B.Y.R. is a cofounder and consultant for Seqstant. All potential conflicts are managed in accordance with institutional policy. K.R. is a shareholder, and Chief Executive Officer of Lyra Labs. This work uses Claude models and was partially funded by Anthropic through their AI for Science program through credits awarded to L.L. This work uses GPT models and was partially funded by OpenAI through credits awarded to L.L. The authors declare that they have no additional competing interests beyond those disclosed above.

\bibliographystyle{unsrt}
\bibliography{reference-full}

\newpage

\appendix
\renewcommand{\thesection}{S\arabic{section}}
\section{Supplementary Information}\label{sec11}

\renewcommand{\thefigure}{S\arabic{figure}}
\renewcommand{\thetable}{S\arabic{table}}

\setcounter{figure}{0}
\setcounter{table}{0}



\renewcommand{\figurename}{Supplementary Figure}
\setcounter{figure}{0}
\section*{Supplementary Figures}

\begin{figure*}[htb]
    \centering
    \includegraphics[width=\linewidth]{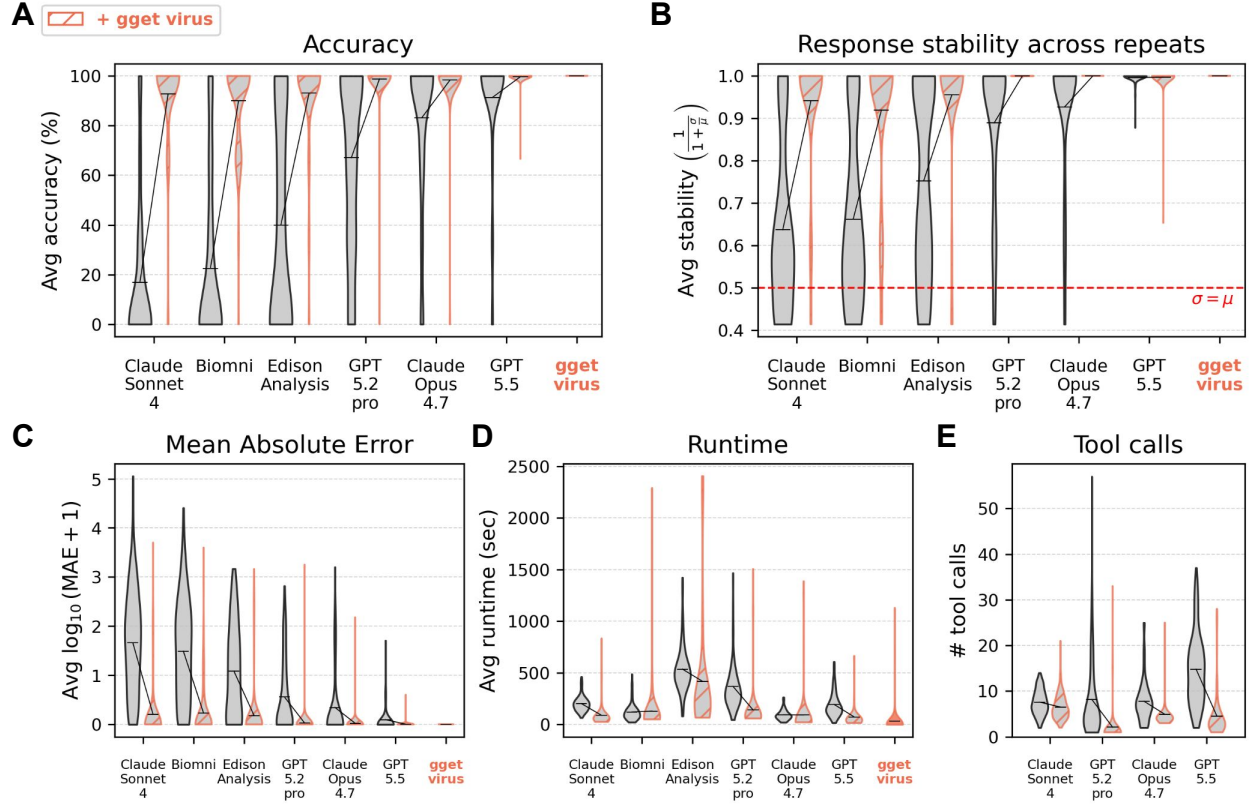}
    \caption{\footnotesize
    \textbf{Per-query distributions of agent performance metrics.} Violin plots of the results shown in \Cref{fig:agents} showing the distribution across all benchmark queries (each executed three times). Black horizontal bars indicate group means, and connecting lines link the corresponding without-\textit{gget virus} and with-\textit{gget virus} conditions for each agent. VirBench results are shown for direct querying of \textit{gget virus}, and the Biomni \cite{biomni}, Edison Analysis \cite{edison_analysis}, Claude Sonnet 4 (with web search and code execution tools), and GPT-5.2-pro (with web search and code execution tools) agents, with and without being prompted to use \textit{gget virus}.
    \textbf{(A)} Accuracy, defined per query as the fraction of runs (out of three) that exactly match the expected sequence count.
    \textbf{(B)} Stability scores per query, computed as $1/(1+\sigma/\mu)$, where $\mu$ and $\sigma$ denote the mean and standard deviation of retrieved counts across runs. The dashed line marks the condition $\sigma = \mu$, corresponding to major instability across repeated runs.
    \textbf{(C)} Error magnitude per query, measured as $\log_{10}(\mathrm{MAE}+1)$, where MAE denotes mean absolute error between retrieved and expected counts across runs. 
    \textbf{(D)} Runtime per query (seconds), averaged across the three repeat runs. 
    \textbf{(E)} Number of agent tool calls, averaged across the three repeat runs.
    }
    \label{fig:supp_agents}
\end{figure*}

\clearpage

\begin{figure*}[htb]
    \centering
    \includegraphics[width=\linewidth]{images/comparison_heatmaps.pdf}
    \caption{\footnotesize
    \textbf{Agent performance on VirBench across taxonomies.} n denotes the number of queries covering each taxonomy.
    \textbf{(A)} Mean runtime (in seconds) per query.
    \textbf{(B)} Mean stability per query. Stability is computed as $1/(1+\sigma/\mu)$, where $\mu$ and $\sigma$ denote the mean and standard deviation of retrieved counts across runs.
    \textbf{(C)} Mean error magnitude per query, measured as $\log_{10}(\mathrm{MAE}+1)$, where MAE denotes mean absolute error between retrieved and expected counts across runs.
    }
    \label{fig:comparison_heatmaps}
\end{figure*}

\clearpage

\begin{figure*}[htb]
    \centering
    \includegraphics[width=0.94\linewidth]{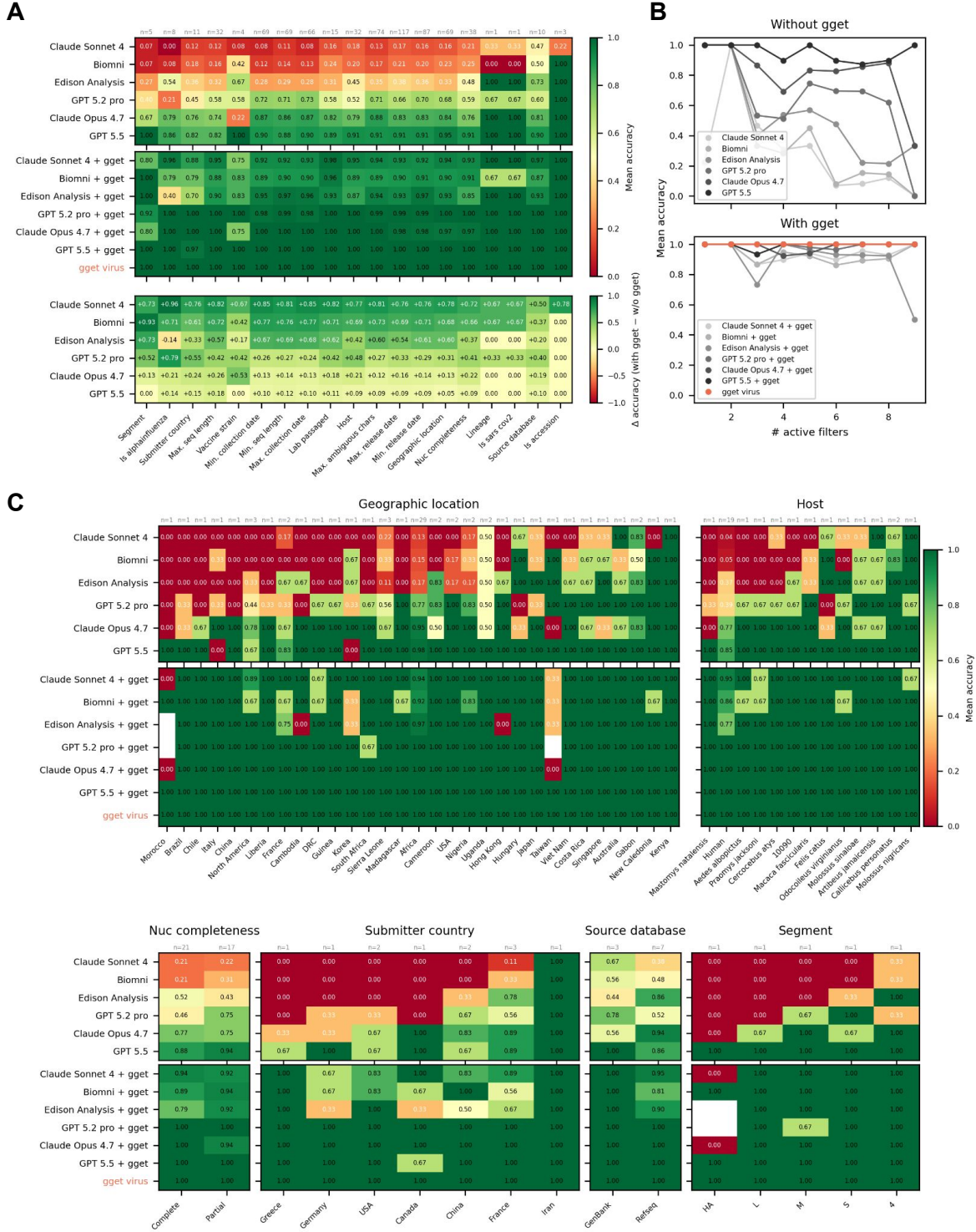}
    \caption{\footnotesize
    \textbf{Influence of filters on agent VirBench performance.}
    \textbf{(A)} Mean accuracy when each filter condition is present across agents, shown separately for agents without and with \textit{gget virus}, and for the standalone \textit{gget virus} baseline. The lower panel shows the change in accuracy relative to the non-\textit{gget} setting.
    \textbf{(B)} Mean accuracy as a function of the number of active filters per query without (top) and with \textit{gget virus} (bottom).
    \textbf{(C)} Mean accuracy stratified by individual filter categories, including geographic location, host, nucleotide completeness, submitter country, source database, and genomic segment.
    }
    \label{fig:filter_heatmaps}
\end{figure*}

\begin{figure*}[htb]
    \centering
    \includegraphics[height=0.75\textheight,width=0.85\linewidth,keepaspectratio]{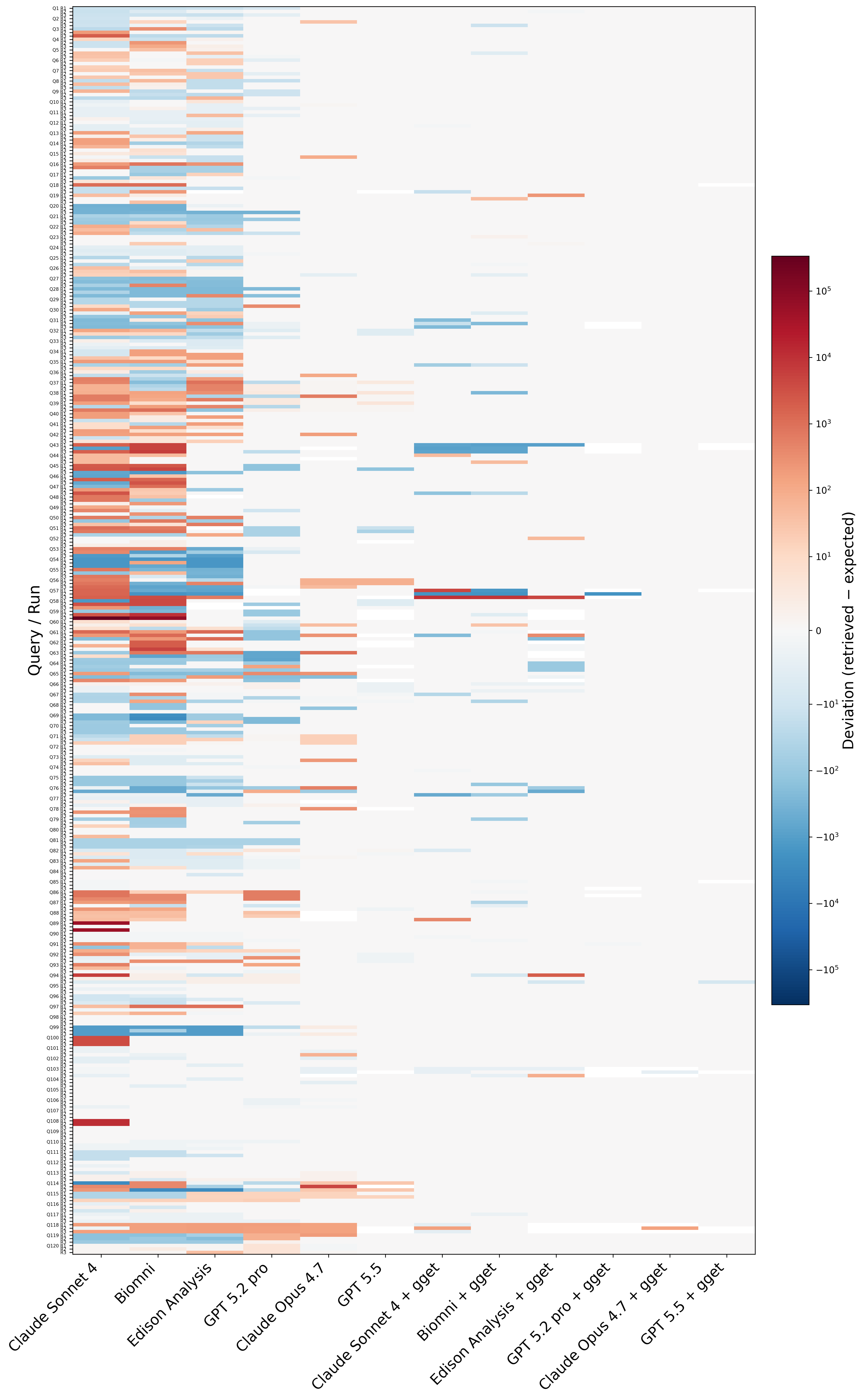}
    \caption{\footnotesize
    \textbf{Deviation between expected and returned VirBench benchmarking results.} Deviations (expected - returned) are shown for each technology for each individual run across 120 queries, each repeated three times.
    }
    \label{fig:all_deviations}
\end{figure*}

\begin{figure*}[htbp]
    \centering
    \includegraphics[width=\linewidth]{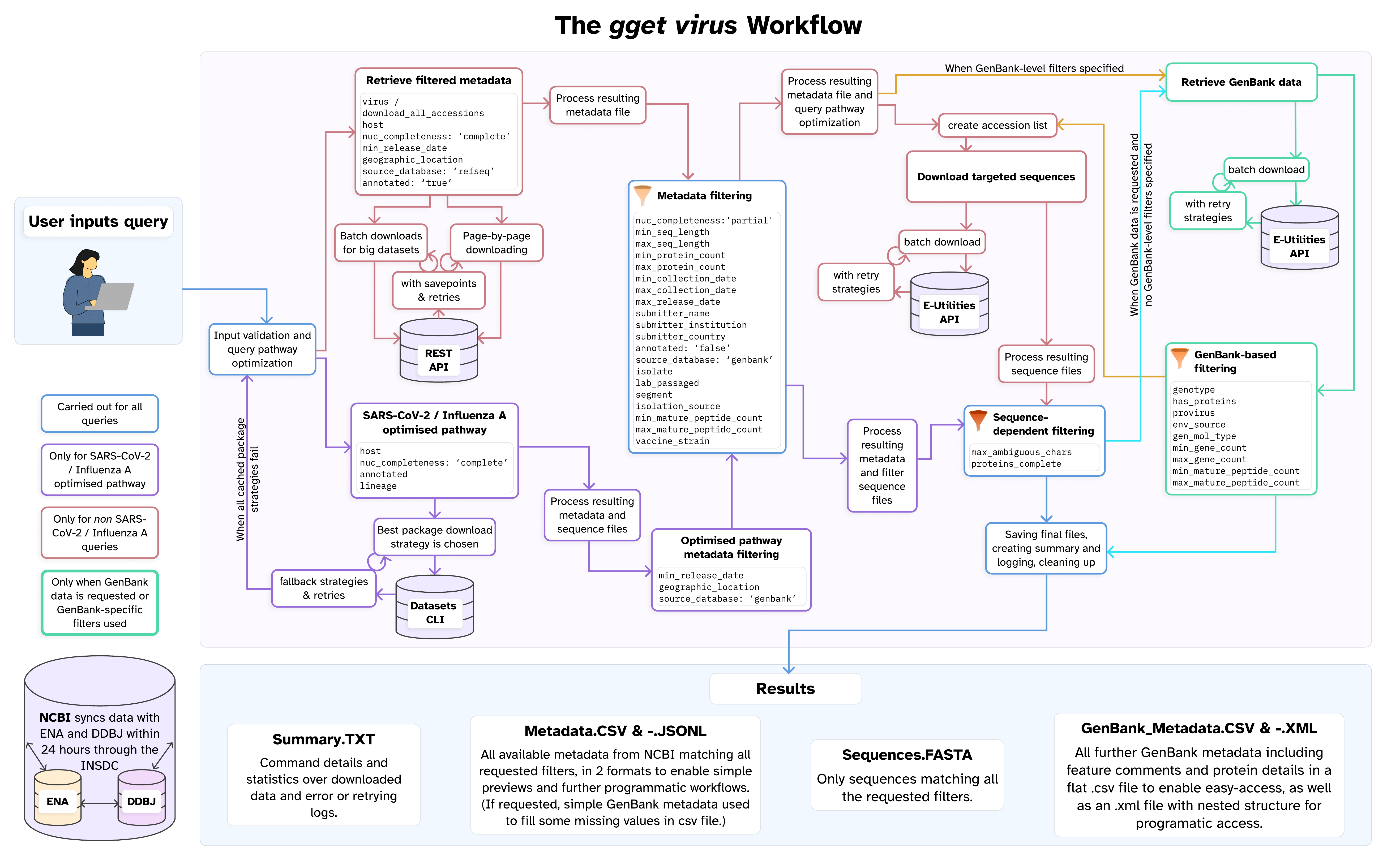} \\
    \caption{\footnotesize
       \textbf{Technical overview of the \textit{gget virus} workflow}. This diagram depicts the internal processing steps triggered by a user query and illustrates how metadata retrieval, layered filtering, sequence download, and optional GenBank enrichment interact within the overall pipeline.
    }
    \label{fig:overview2}
\end{figure*}

\begin{figure*}[p] 
    \centering
    
    \begin{minipage}[b]{0.48\textwidth}
        \centering
        \begin{tikzpicture}
            \node[anchor=south west, inner sep=0] (image) at (0,0) {
                \includegraphics[height=1\textheight, width=\linewidth, keepaspectratio]{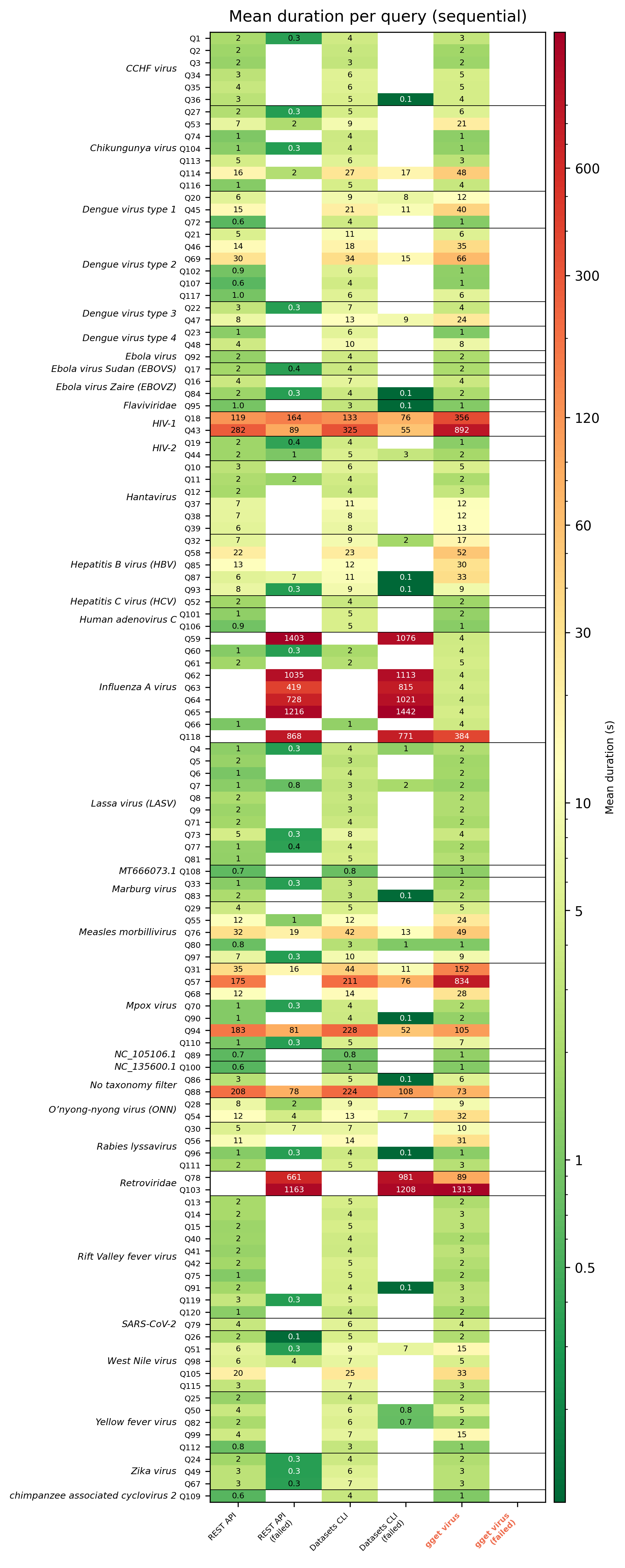}
            };
            \node[anchor=north west, font=\bfseries, xshift=-5pt, yshift=-5pt] at (image.north west) {A};
        \end{tikzpicture}
    \end{minipage}
    \hfill
    \begin{minipage}[b]{0.48\textwidth}
        \centering
        \begin{tikzpicture}
            \node[anchor=south west, inner sep=0] (image) at (0,0) {
                \includegraphics[height=1\textheight, width=\linewidth, keepaspectratio]{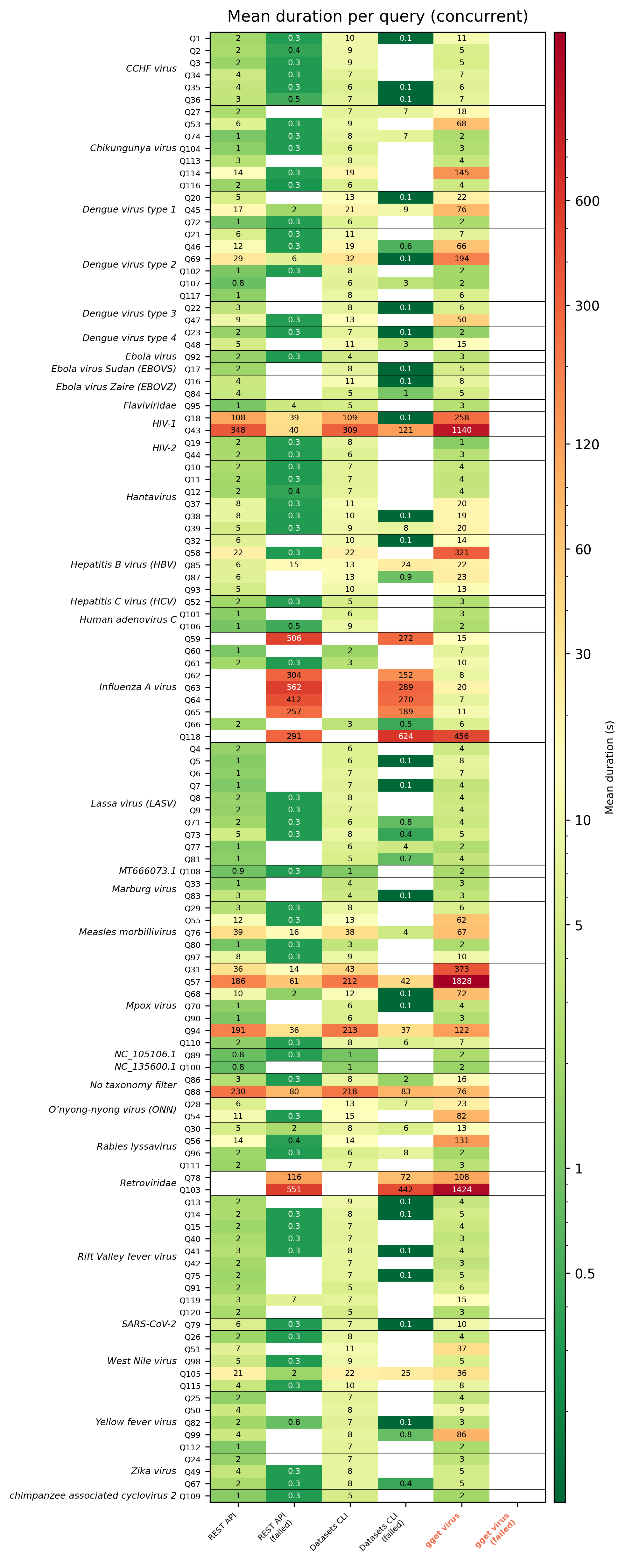}
            };
            \node[anchor=north west, font=\bfseries, xshift=-5pt, yshift=-5pt] at (image.north west) {B};
        \end{tikzpicture}
    \end{minipage}
    \caption{\footnotesize \textbf{Runtime per query for concurrent and sequential benchmarking of programmatic viral data retrieval methods based on 120 queries every hour over 24 hours.} The heatmap shows the mean runtime per query. Failed queries are plotted separately from successful queries. White cells represent no runs in that category (e.g., when failed runs are white, the query must have terminated successfully over all 24 rounds with that method). Queries are grouped by virus name.
    \textbf{(A)} Sequential (one query at a time).
    \textbf{(B)} Concurrent (three simultaneous queries at a time). An overview of this plot grouped by virus names is shown in \Cref{fig:benchmark-methods} E.
    }
    \label{fig:runtime_heatmaps}
\end{figure*}

\begin{figure*}[htb] 
    \centering
    \includegraphics[width=\linewidth]{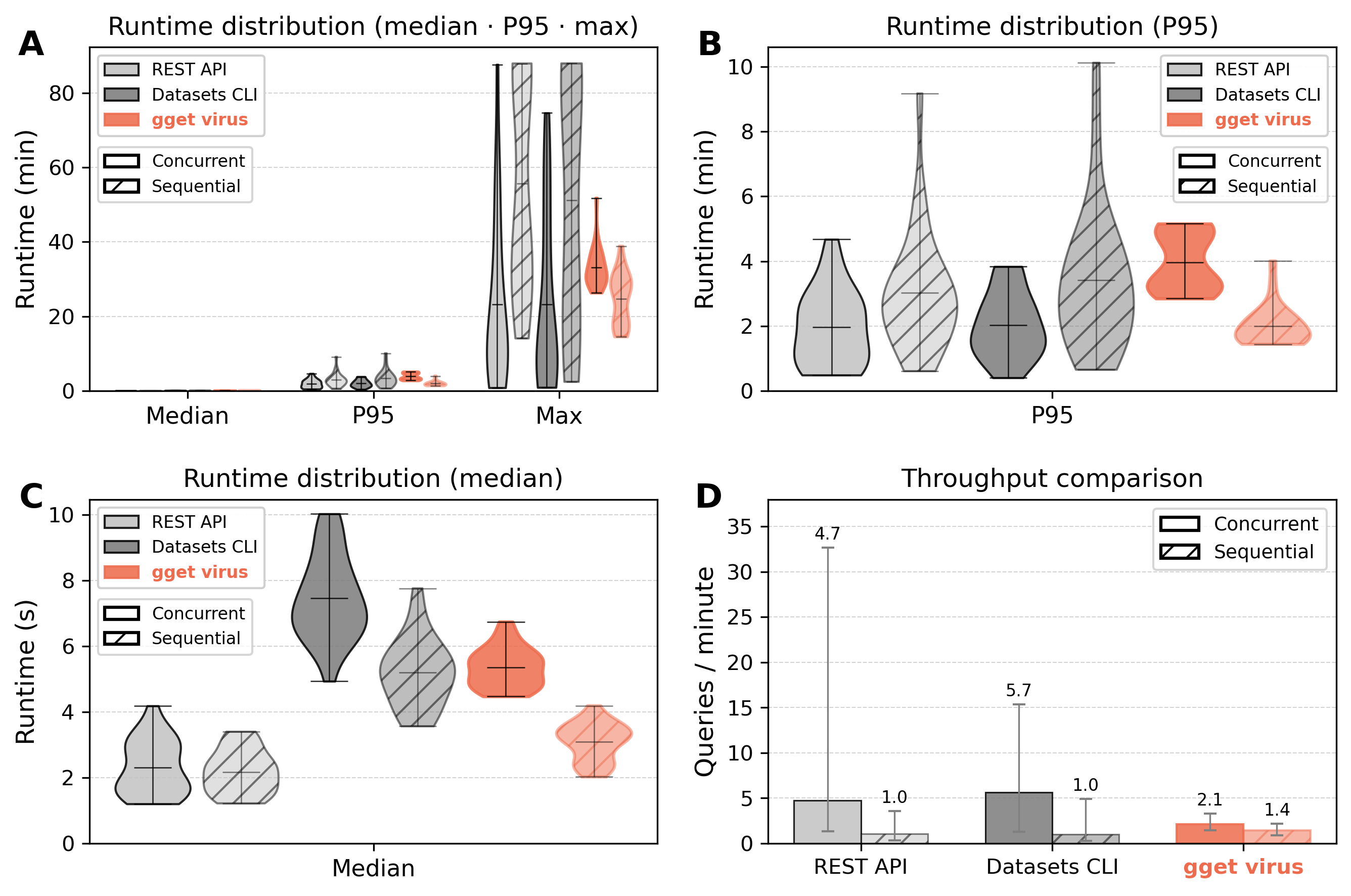}
    \caption{\footnotesize \textbf{Runtime distribution and throughput comparison of the sequential and concurrent (three simultaneous queries at a time) benchmarking of programmatic viral data retrieval methods based on 120 queries every hour over 24 hours.} 
    \textbf{(A-C)} Runtime distribution for all methods, shown as violin plots for the median, 95th-percentile, and maximum query durations over all query repetitions. Mean values for each violin plot are marked. 95th percentile and median distributions are plotted separately with differing y-axes.
    \textbf{(D)} Number of queries fulfilled per minute. The bars show the mean value over all runs, while the error bars denote the minimum and maximum throughput for each method.
    }
    \label{fig:runtime-dist-throughput}
\end{figure*}

\begin{figure*}
\centering
    \includegraphics[height=0.78\textheight, width=\linewidth, keepaspectratio]{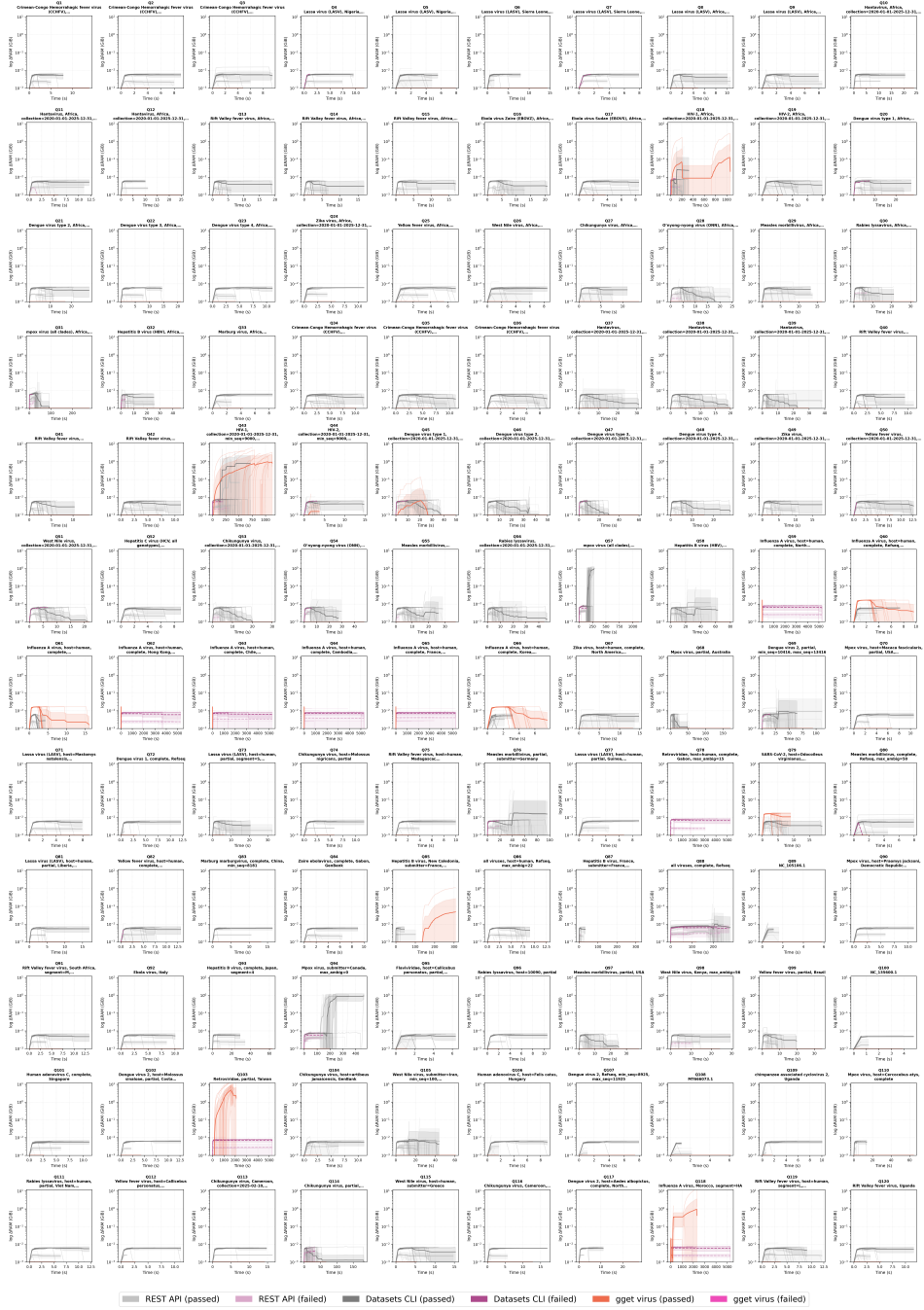}
    \caption{\footnotesize \textbf{RAM usage per query for all 120 VirBench queries run sequentially every hour over 24 hours.} The median usage per method across queries is shown as a thick line, with individual trajectories overlaid as fainter lines; the upper and lower bounds are indicated by a shaded region.}
    \label{fig:ram-usage}
\end{figure*}

\clearpage

\renewcommand{\tablename}{Supplementary Table}
\setcounter{table}{0}
\section*{Supplementary tables}

\begin{table*}[ht]
\centering
\caption{Overview of major NCBI APIs and data access pathways.}
\label{tab:ncbi_apis}
\begin{tabularx}{\linewidth}{
>{\raggedright\arraybackslash}p{0.13\linewidth}
>{\raggedright\arraybackslash}p{0.16\linewidth}
>{\raggedright\arraybackslash}p{0.18\linewidth}
>{\raggedright\arraybackslash}X
>{\raggedright\arraybackslash}X
}
\textbf{Resource} 
& \textbf{Primary Use Case} 
& \textbf{URL} 
& \textbf{Strengths} 
& \textbf{Weaknesses} \\ \hline

E-utilities (Entrez programming utilities) 
& Structured interface to the Entrez system 
& \url{https://www.ncbi.nlm.nih.gov/books/NBK25501/} 
& Flexible; widely supported; works across many databases 
& Requires URL creations; manual batching \\ \hline

NCBI Virus REST API 
& General-purpose queries across NCBI databases  
& \url{https://www.ncbi.nlm.nih.gov/datasets/docs/v2/api/rest-api/} 
& Modern API; supports some server-side filtering; clean JSON output
& Requires multi-step workflows; limited filtering (\Cref{tab:feature-comp}); manual pagination  \\ \hline

NCBI Datasets CLI 
& Structured bulk downloads (genomes, viruses, taxonomy) 
& \url{https://www.ncbi.nlm.nih.gov/datasets/docs/v2/} 
& Supports bulk retrieval of data; includes cached datasets
& Less flexible filtering; most metadata fields not exposed (\Cref{tab:feature-comp})\\ \hline

NCBI FTP (Bulk downloads) 
& Full dataset access (GenBank, RefSeq, virus datasets) 
& \url{https://ftp.ncbi.nlm.nih.gov/} 
& Complete datasets; efficient for large-scale analysis 
& No server-side filtering; requires local parsing and processing \\ \hline

SRA Toolkit
& Raw sequencing reads (SRA) 
& \url{https://github.com/ncbi/sra-tools/wiki} 
& Direct access to raw sequencing data; standardized tools 
& Not suited for assembled genomes or metadata-heavy filtering \\ \hline

\end{tabularx}
\end{table*}

\vspace{2cm}

\begin{table*}[ht]
\centering
\caption{Source code and documentation hyperlinks.}
\label{tab:links}
\begin{tabularx}{\linewidth}{p{0.3\linewidth} p{0.7\linewidth}}
\textbf{Description} & \textbf {Link}  \\  \hline
\textit{gget virus} source code                                          & \url{https://github.com/pachterlab/gget/blob/main/gget/gget\_virus.py}  \\  \hline
English tool documentation                                                        & \url{https://pachterlab.github.io/gget/en/virus.html} \\  \hline
Spanish tool documentation                                                        & \url{https://pachterlab.github.io/gget/es/virus.html} \\  \hline  
Google Colab tutorials                                                       & \url{https://github.com/pachterlab/gget_examples/blob/main/gget_virus} \\  \hline
VirBench analysis                                                      & \url{https://github.com/lauraluebbert/VirBench} \\  \hline
\end{tabularx}
\end{table*}

\begin{table*}[htbp]
\centering
\caption{
Statistical comparisons of \textit{gget virus} against baseline methods. All tests are two-sided Mann--Whitney $U$ tests. All $p$-values are corrected using Bonferroni adjustment over $k=8$ tests ($\alpha_{\mathrm{adj}} = 0.00625$).}
\label{tab:stat_tests}
\small
\resizebox{\textwidth}{!}{%
\begin{tabular}{@{}ll rr rr rc@{}}
\toprule
\textbf{Metric} & \textbf{Comparison} &
  \makecell[c]{\textbf{\textit{n}}\\\textbf{\textit{gget virus}}} &
  \makecell[c]{\textbf{\textit{n}}\\\textbf{Baseline}} &
  \makecell[c]{\textbf{Median}\\\textbf{\textit{gget virus}}} &
  \makecell[c]{\textbf{Median}\\\textbf{Baseline}} &
  \makecell[c]{\textbf{$U$}\\\textbf{statistic}} &
  \textbf{\textit{p}-value} \\
\midrule
\multirow{2}{*}{\makecell[l]{Peak Storage\\(MB\,/\,query)}}
  & vs.\ Datasets REST API      & $2{,}880$ & $2{,}543$ & $0.4$  & $2.2$ & $2{,}584{,}049$  & $< 10^{-16}$ \\
  & vs.\ Datasets CLI  & $2{,}880$ & $2{,}598$ & $0.4$  & $2.6$ & $2{,}615{,}604$  & $< 10^{-16}$ \\
\addlinespace
\multirow{2}{*}{\makecell[l]{Metadata fields\\(enriched CSV)}}
  & vs.\ Datasets REST API (JSON)      & $2{,}787$ & $2{,}519$ & $41$  & $25$ & $7{,}020{,}453$  & $< 10^{-16}$ \\
  & vs.\ Datasets CLI (JSON)  & $2{,}787$ & $2{,}576$ & $41$ & $25$ & $7{,}179{,}312$  & $< 10^{-16}$ \\
\addlinespace
\multirow{2}{*}{\makecell[l]{Metadata fields\\(enriched XML)}}
  & vs.\ Datasets REST API (JSON)      & $2{,}787$ & $2{,}519$ & $51$ & $25$ & $7{,}020{,}453$ & $< 10^{-16}$ \\
  & vs.\ Datasets CLI (JSON)  & $2{,}787$ & $2{,}576$ & $51$ & $25$ & $7{,}179{,}312$  & $< 10^{-16}$ \\
\addlinespace
\multirow{2}{*}{\makecell[l]{Peak RAM\\(GB)}}
  & vs.\ Datasets REST API      & $2{,}880$ & $2{,}543$ & $18.9$ & $18.3$ & $3{,}591{,}561$  & $0.22$ \\
  & vs.\ Datasets CLI  & $2{,}880$ & $2{,}598$ & $18.9$ & $19$ & $3{,}671{,}200$  & $0.23$ \\
\bottomrule
\end{tabular}
}

\end{table*}


\begin{table*}[p]
\centering
\caption{\textbf{Data retrieval process and size comparison.} \newline The steps involved in viral sequence data retrieval are described for four different methods, along with the total size and number of records of data downloaded. Some methods retrieve broader datasets that require additional manual filtering by the user.}
\label{tab:process_comparison}
\scriptsize
\setlength{\tabcolsep}{3pt}
\renewcommand{\arraystretch}{0.9}
\begin{tabularx}{\linewidth}{p{0.09\linewidth}|p{0.28\linewidth} X X X}
\hline
\textbf{Feature} & \textbf{NCBI Web} & \textbf{NCBI Datasets REST API} & \textbf{Datasets CLI} & \textbf{gget virus} \\ \hline
\multicolumn{5}{p{0.99\linewidth}}{\textit{Example 1: Retrieval of 2025 SARS-CoV-2 sequences with the surface glycoprotein}} \\ 
\textbf{Process} &
Visit this web page: \url{https://www.ncbi.nlm.nih.gov/labs/virus/vssi/\#/} \newline 
Insert Virus name: "Severe acute respiratory syndrome coronavirus 2" or the Taxon ID "2697049" or click on the "SARS-CoV-2" button. \newline 
Using the filter categories on the left-hand side of the next page, choose "Dates" and choose "Release Date" then enter the dates (or use the calendar option) to enter from "01/02/2025" to "01/01/2026". Note that the web page accepts MM/DD/YYYY format and it changes the start date by one day after pressing "submit", and is not inclusive on the maximum date, therefore we proactively insert the 2nd of January as the start date, and 1st of January 2026 as the end date. \newline 
Next, choose the "Genome Organization" category and in "Has Proteins" insert "surface glycoprotein". To download the sequences, click on "Download All Results" at the top of the table. Then go through 3 pages of details, choosing "Nucleotide" under the Sequence Data column, then "Download All Records" choosing the default FASTA file description format and pressing "Download". \newline 
Downloading the metadata is a separate task; one has to again click on "Download All Results", choose "CSV format" under the "Results Table" column, then "Download All Records", followed by "Select All" columns with accession version, and pressing "Download". To get the programmatic metadata version "XML format", this last part is repeated. & 
With the following curl command, while using all the allowed relevant filters, we can download the SARS-CoV-2 data released since 2025-01-01:
\newline \newline
\texttt{curl -X GET "https://api.ncbi.nlm. nih.gov/datasets/v2 /virus/taxon/2697049 /genome/download? released\_since= 2025-01-01\&include\_ sequence=GENOME \&aux\_report=DATASET\_ REPORT" -H 'accept: application/zip' -o sars\_cov\_2.zip}  \newline
\newline
The process for this method is not exactly comparable to the others, as there is a much larger learning curve and experience required to set up a command like this in comparison to the other methods described here. & 
\texttt{pip install datasets} \newline
\texttt{datasets download virus genome taxon SARS-CoV-2 --filename sars\_cov\_2.zip} \newline
\newline
This is the only cached file available for download that contains all the data we require; however, it also includes substantially more data, as filtering to the desired level is not possible using the \textit{Datasets CLI} alone. & 
\texttt{uv pip install gget} \newline
\texttt{gget virus "SARS-CoV-2" --min\_release\_date 2025-01-01 --max\_release\_date 2025-12-31 --has\_proteins "surface glycoprotein"} \\
\textbf{\# Records} & \textbf{115,987} & \textbf{171,371} & \textbf{9,192,765} & \textbf{115,987} \\
\textbf{Data size} & \textbf{3.59 GB} & \textbf{5.3 GB} & \textbf{284 GB} & \textbf{3.83 GB} \\ \hline
\end{tabularx}
\end{table*}

\begin{table*}[p]
\ContinuedFloat
\centering
\caption{\textbf{Data retrieval process and size comparison.} \textit{(continued)}}
\scriptsize
\setlength{\tabcolsep}{3pt}
\renewcommand{\arraystretch}{0.9}
\begin{tabularx}{\linewidth}{p{0.09\linewidth}|p{0.28\linewidth} X X X}
\hline
\textbf{Feature} & \textbf{NCBI Web} & \textbf{NCBI Datasets REST API} & \textbf{Datasets CLI} & \textbf{gget virus} \\ \hline
\multicolumn{5}{p{0.99\linewidth}}{\textit{Example 2: Retrieval of all human viral sequences from Switzerland released before January 1st, 2000}} \\
\textbf{Process} &
Similar to above, choose the taxon ID: "10239" or "All Viruses" from the first page. Under "Location and Source", type: "Switzerland" under "Geographic Region" and choose "Human" under "Host". When setting the “Release Date” in the “Dates” section, the interface requires a start date even if we only want to filter by end date. Therefore, when no lower bound is intended, a very early date (e.g., January 1, 1900) must be selected in order to set the desired maximum release date, such as January 1, 2000. The resulting records can be downloaded as described above. &
\texttt{curl -X GET "https://api.ncbi.nlm. nih.gov/datasets/v2 /virus/taxon/10239 /genome/download? host=9606\&geo\_location= Switzerland\&include\_ sequence=GENOME\& aux\_report=DATASET\_REPORT" -H 'accept: application/zip' -o all\_swiss\_viruses.zip} &
\texttt{pip install datasets} \newline
\texttt{datasets download virus genome taxon 10239 --host human --geo-location Switzerland --filename all\_swiss\_viruses.zip} & 
\texttt{uv pip install gget} \newline
\texttt{gget virus --download\_all\_accessions --host human --max\_release\_date 2000-01-01 --geographic\_location Switzerland} \\
\textbf{\# Records} & \textbf{47} & \textbf{186,208} $\ast$ & \textbf{186,208} $\ast$ & \textbf{47} \\
\textbf{Data size} & \textbf{0.12 MB} & \textbf{5,059 MB} $\ast$ & \textbf{5,059 MB} $\ast$ & \textbf{0.22 MB} \\ \hline
\end{tabularx}
\caption*{\footnotesize $\ast$ This is an estimated number as the query fails as of writing this paper (printed reason is 'Internal error'). We have been in contact with NCBI, who have acknowledged the error and are working to resolve it shortly. In the meantime, experienced users can work around the problem by using POST requests. \textit{gget virus} includes fallback strategies that circumvent these points of failure.}
\end{table*}

\begin{table*}[p]
\ContinuedFloat
\centering
\caption{\textbf{Data retrieval process and size comparison.} \textit{(continued)}}
\scriptsize
\setlength{\tabcolsep}{3pt}
\renewcommand{\arraystretch}{0.9}
\begin{tabularx}{\linewidth}{p{0.09\linewidth}|p{0.28\linewidth} X X X}
\hline
\textbf{Feature} & \textbf{NCBI Web} & \textbf{NCBI Datasets REST API} & \textbf{Datasets CLI} & \textbf{gget virus} \\ \hline
\multicolumn{5}{p{0.99\linewidth}}{\textit{Example 3: Retrieval of proviral HIV1 sequences released before August 31st, 2025, with a minimum sequence length of 1000 bases, including the ”gag polyprotein”, as shown in \Cref{fig:overview}.}} \\
\textbf{Process} &
Similar to above, one has to search for "Human immunodeficiency virus 1" or the Taxon ID: "11676". The maximum release date needs to be chosen like above by also choosing a minimum release date. Under "Sequence Quality" and "Sequence Length", the minimum can be set to 1000. Under "Genome Organization," the "gag polyprotein" needs to be named in the "Has Proteins" section, and under "Provirus/Integrated", the option "Only" should be selected. The resulting records are downloaded as described above. & 
\texttt{curl -X GET "https://api.ncbi.nlm. nih.gov/datasets/v2/ virus/taxon/11676 /genome/download?\& include\_sequence= GENOME\&aux\_report= DATASET\_REPORT" -H 'accept: application/zip' -o hiv1.zip} & 
\texttt{pip install datasets} \newline
\texttt{datasets download virus genome taxon 11676 --filename hiv1.zip} &
\texttt{uv pip install gget} \newline
\texttt{gget virus hiv1 --max\_release\_date 2025-08-31 --min\_seq\_length 1000 --has\_proteins "gag polyprotein" --provirus true} \\
\textbf{\# Records} & \textbf{274} $\dagger$ & \textbf{1,333,382} & \textbf{1,333,382} & \textbf{309} $\dagger$ \\
\textbf{Data size} & \textbf{1.35 MB} & \textbf{5,229 MB} & \textbf{5,229 MB} & \textbf{2.34 MB} \\ \hline
\end{tabularx}
\caption*{\footnotesize $\dagger$ \textit{gget virus} retrieves more records than NCBI Web because the protein filter on the website matches only exact strings, omitting records in which the protein name differs in capitalization, such as "Gag protein" rather than "gag protein."}
\end{table*}

\end{document}